\documentclass[galaxies,article,accept,pdftex,moreauthors]{Definitions/mdpi} 

\firstpage{1} 
\pubvolume{1}
\issuenum{1}
\articlenumber{0}
\pubyear{2024}
\copyrightyear{2024}
\datereceived{ } 
\daterevised{ } 
\dateaccepted{ } 
\datepublished{ } 
\hreflink{https://doi.org/} 

\Title{Scaling relations of early-type galaxies in MOND}

\TitleCitation{Shape and morphology of early-type galaxies in MOND}

\Author{Robin Eappen $^{1}$\orcidA{} and Pavel Kroupa $^{1,2}$\orcidB{}}

\AuthorNames{Robin Eappen and Pavel Kroupa}

\AuthorCitation{Eappen, R.; Kroupa, P.}

\address{%
$^{1}$ \quad Helmholtz-Institut für Strahlen- und Kernphysik (HISKP), Universität Bonn, Nussallee 14–16, 53115 Bonn, Germany\\
$^{2}$ \quad Charles University in Prague, Faculty of Mathematics and Physics, Astronomical Institute, V Holešovickách 2, CZ-180 00 Praha 8, Czech Republic}

\corres{Correspondence: robin.eappen@gmail.com (R.E.), pkroupa@uni-bonn.de(P.K.)}

\abstract{We investigate the shape and morphology of early-type galaxies (ETGs) within the framework of Modified Newtonian Dynamics (MOND). Building on our previous studies, which demonstrated that the monolithic collapse of primordial gas clouds in MOND produces galaxies (noted throughout as 'model relics' in the context of this work) with short star formation timescales and a downsizing effect as observationally found, we present new analyses on the resulting structural and morphological properties of these systems. Initially, the monolithically formed galaxies display disk-like structures. In this study, we further analyze the transformations that occur when these galaxies merge, observing that the resulting systems (noted throughout as 'merged galaxies' in the context of this work) take on elliptical-like shapes, with the $(V_{\mathrm{rot}} / V_{\sigma})$–ellipticity relations closely matching observational data across various projections. We extend this analysis by examining the isophotal shapes and rotational parameter $(\lambda_{R})$ of both individual relics and merged galaxies. The results indicate that ETGs may originate in pairs in dense environments, with mergers subsequently producing elliptical structures that align well with observed kinematic and morphological characteristics. Finally, we compare both the model relics and merged galaxies with the fundamental plane and Kormendy relation of observed ETGs, finding close agreement. Together, these findings suggest that MOND provides a viable physical framework for the rapid formation and morphological evolution of ETGs.}


\begin{document}

\section{Introduction}

The formation and evolution of early-type galaxies (ETGs) remain central questions in cosmology. ETGs, which are predominantly elliptical in morphology, exhibit a wide range of structural and dynamical characteristics, reflecting their diverse formation histories. Observationally, ETGs display shapes ranging from nearly spherical systems to highly flattened, triaxial configurations, with their ellipticity often correlating with mass and environment (\citealt{2009ApJS..182..216K}). For instance, more massive ETGs tend to be rounder, while lower-mass ETGs frequently retain disc-like features \citep{2011MNRAS.414..888E}. Large surveys such as ATLAS3D (\citealt{2011MNRAS.413..813C}) have further refined our understanding, showing that ETGs can be classified as fast or slow rotators based on kinematic properties.

The environment also influences ETG morphology, with denser cluster environments favoring rounder, more massive ETGs, while field ETGs often exhibit disc-like features, suggesting quieter evolutionary paths (\citealt{1980ApJ...236..351D, 1993ApJ...407..489W}). Observations of high-redshift galaxies, reveal compact, spheroidal shapes at \( z > 2 \), indicating a transition to the more diverse morphologies observed in local ETGs (\citealt{2007MNRAS.382..109T, 2009ApJ...700..221K}). These empirical findings provide critical benchmarks for evaluating theoretical models of ETG formation and evolution, underscoring the importance of understanding their shape and morphology in various cosmological contexts.

Recent observations have added further detail to our understanding of ETGs and high-redshift galaxies. Massive, luminous galaxies observed at redshifts around \( z \approx 15 \) (\citealt{2022ApJ...939L..31H,2023MNRAS.518.6011D, 2023ApJS..265....5H, 2024ApJ...960...56H}) point to active star formation at very early epochs. These findings, coupled with studies of the fundamental plane (\citealt{1987ApJ...313...59D}), ellipticity relations (\citealt{1978MNRAS.183..501B}), and the Kormendy relation (\citealt{1977ApJ...218..333K}), provide essential benchmarks for evaluating theoretical models. The standard cosmological model, $\Lambda$CDM (Dark Energy Cold Dark Matter), has successfully explained many of these observed correlations through advanced hydrodynamical simulations, offering significant insights into the large-scale structure and galaxy evolution (see \citealt{2005Natur.435..629S,  2015MNRAS.446..521S, Pillepich_2018} for more details on $\Lambda$CDM results on ETG formation).

In parallel, alternative frameworks such as Modified Newtonian Dynamics (MOND) provide a complementary approach to understanding galaxy dynamics and structure without invoking dark matter (\citealt{2012PASA...29..395K, 2015CaJPh..93..169K, 2022Symm...14.1331B, 2022PASP..134l1001M, 2023arXiv230911552K}). This paper focuses on investigating ETG formation within the MOND paradigm, emphasizing the role of shape and morphology as diagnostic tools for evaluating galaxy formation scenarios. Building on prior work that examined monolithic collapse within MOND (\citealt{2022MNRAS.516.1081E}), we extend this analysis by incorporating merger-driven formation scenarios. 

Our study specifically investigates whether MOND-based simulations can reproduce key structural properties of ETGs, including their mass-size relation, velocity dispersion profiles, and ellipticity distributions. By comparing these results with empirical correlations, we aim to test the viability of MOND in explaining the observed diversity of ETG shapes and morphologies. Furthermore, this work contributes to MOND-based cosmological models, such as the $\nu$HDM model (\citealt{2013ApJ...772...10K, 2023MNRAS.523..453W}), which accommodate the rapid formation of massive galaxies at high redshifts (\( z > 10 \), \citealt{2024Univ...10...48M}).

Section \ref{sec:two} outlines the MOND theory and describes the numerical hydrodynamical code employed. Section \ref{sec:models1} details the simulated formation scenarios. Section \ref{sec:results} presents the results, and Section \ref{sec:conclusion} offers concluding remarks. Through this analysis, we aim to provide new insights into ETG shapes and morphologies within MOND, enriching the broader discussion of galaxy formation models.

\section{MOND and Phantom of RAMSES (PoR)}
\label{sec:two}

By combining observational constraints from both galactic dynamics and the Solar System, \cite{1983ApJ...270..365M} proposed a modification to the theory of gravitation at low accelerations (\citealt{2012LRR....15...10F,2025arXiv250117006F}). MOND is formulated through a Lagrangian which, when extremized, produces a generalized Milgromian Poisson equation. Two such Lagrangians are AQUAL \citep{1984ApJ...286....7B} and QUMOND \citep{2010MNRAS.403..886M}. QUMOND is employed in this study as it provides a quasi-linear formulation of MOND, facilitating the numerical modeling of gravitational interactions without the need for dark matter. Its mathematical structure simplifies the implementation of MOND in hydrodynamical simulations by allowing the use of Poisson’s equation, making it computationally more tractable than the fully nonlinear AQUAL formulation. Additionally, QUMOND has been widely adopted in previous studies of galaxy formation within Milgromian dynamics, ensuring consistency and comparability of results within this framework. The generalized Poisson equation is,

\begin{equation} 
\Delta \Phi(\Vec{x}) = 4\pi G\rho_{\rm b} (\Vec{x}) + \Vec{\nabla} \cdot [\tilde{v}(\left\lvert\vec{\nabla} \phi \right\rvert/a_0)\Vec{\nabla} \phi (\Vec{x})],
\label{eq:mil1}
\end{equation}
or,
\begin{equation} 
\Delta \Phi(\Vec{x}) = 4\pi G [\rho_{\rm b} (\Vec{x}) + \rho_{\rm ph}(\Vec{x})],
\label{mil2}
\end{equation}
Here, $\rho_{\rm b}(\vec{x})$ represents the baryonic density at location $\Vec{x}$, and $\phi(\vec{x})$ is the Newtonian potential satisfying the standard Poisson equation $\Delta \phi(\vec{x}) = 4\pi G \rho_{\rm b}(\vec{x})$. Milgrom's constant is defined as $a_0 \approx 1.2 \times 10^{-10} \ {\rm m}{\rm s}^{-2} \approx 3.8 \ {\rm pc} \ {\rm Myr}^{-2}$. The term $\rho_{\rm ph}(\vec{x})$, or phantom dark matter (PDM) density, is a mathematical construct arising from the non-linearity of the generalized Poisson equation, rather than being an actual matter density distribution. The total gravitational potential, $\Phi(\vec{x})$, yields the accelerations via $\vec{a} = -\vec{\nabla} \Phi$, with $\tilde{v}(y)$ serving as a transition function characterizing the theory \citep{2008arXiv0801.3133M,2010MNRAS.403..886M,2014SchpJ...931410M,2012LRR....15...10F,2022Symm...14.1331B}. The transition function $\tilde{v}(y)$ has the following asymptotic behavior:

\begin{equation}
\tilde{v}(y) \rightarrow 0 \ \mathrm{for} \ y \gg 1 \ \mathrm{and} \  \tilde{\it v}(y) \rightarrow y^{-1/2} \ \mathrm{for} \ y \ll 1.
\label{mil01}
\end{equation}

This approach involves only linear differential equations and emerges naturally from a Palatini-type formulation of Newtonian gravity, placing it within a broader class of bi-potential theories (QUMOND, \citealt{2010MNRAS.403..886M}). Notably, \citet{1984ApJ...286....7B} observed similarities with certain quark confinement theories when using a different form of the transition function $\tilde{v}$.

MOND has successfully predicted various galactic scaling relations, such as the Baryonic Tully-Fisher Relation (BTFR) \citep{2000ApJ...533L..99M,2005ApJ...632..859M,2012AJ....143...40M,2016ApJ...818..179L} and the Radial Acceleration Relation (RAR) \citep{1990A&ARv...2....1S,2004ApJ...609..652M,2017ApJ...836..152L}.

To solve the equations describing mathematical MOND models and facilitate the study of galaxy formation within a computational domain, numerical simulation techniques are required. One such technique is implemented in the Phantom of RAMSES (PoR) code, developed in Bonn in collaboration with Strasbourg \citep{2015CaJPh..93..232L,2021arXiv210111011N}. PoR is a customized version of the RAMSES code \citep{2002A&A...385..337T}, designed to solve the generalized MOND Poisson equation using adaptive mesh refinement (AMR) and employing a multi-grid and conjugate gradient solver.

Several transition functions are employed in the literature to address the non-linearity of the Poisson equation in MOND. PoR utilizes the function:

\begin{equation}
\tilde{v}(y) = -\frac{1}{2} + \left(\frac{1}{4} + \frac{1}{y}\right)^{\frac{1}{2}},
\label{eq
}
\end{equation}
where $y = \left|\vec{\nabla} \phi\right|/a_{0}$ \citep{2015CaJPh..93..232L}.

The PoR simulation workflow involves first solving the standard Poisson equation to compute the Newtonian potential given only the baryonic matter density $\rho_{\rm b}(\vec{x})$. Next, the phantom dark matter (PDM) density is calculated using a discrete scheme. Subsequently, the Poisson equation is solved again with both the Newtonian and PDM densities to compute the total gravitational potential \citep{2015CaJPh..93..232L}.

The PoR code has been successfully applied in various research projects, including hydrodynamical simulations of Antennae-like galaxies \citep{2016MNRAS.463.3637R}, the Sagittarius satellite galaxy \citep{2017A&A...603A..65T}, and the Local Group to produce the planes of satellites \citep{2018A&A...614A..59B,2018MNRAS.477.4768B,2021Galax...9..100B}. It has also been used in simulations of globular cluster streams \citep{2018A&A...609A..44T}, the simulation of asymmetric tidal tailes of open clusters \citep{2022MNRAS.517.3613K}, the formation of exponential disk galaxies \citep{2020ApJ...890..173W}, the global stability of M33 \citep{2020ApJ...905..135B}, the evolution of globular-cluster systems in ultra-diffuse galaxies due to dynamical friction \citep{2021arXiv210705667B}, and the modeling of polar-ring galaxies using a pre-PoR code \citep{2013MNRAS.432.2846L}. A comprehensive user guide for PoR is available in \citet{2021CaJPh..99..607N}. Structure formation simulations in the $\nu$HDM cosmological model have also been published \citep{2023MNRAS.523..453W}.

\citet{2022MNRAS.516.1081E} utilized PoR to constrain the initial conditions of post-Big-Bang gas clouds necessary for forming ETGs. In our study, we adopt model galaxies (relics from here on) from \citet{2022MNRAS.516.1081E} that form from collpasing post-Big-Bang gas clouds with an initial radii of 500 kpc and examine the mergers formed from these initial conditions. These initial radii of pre-galactic gas clouds of different masses lead to the observationally deduced downsizing (i.e. formation) times of ETGs.

\section{Models}
\label{sec:models1}

In this study, we examine the structural, morphological, and shape features of galaxies that arise from the monolithic collapse of post-Big-Bang gas clouds, as well as the properties of galaxies resulting from the merger of these collapsed systems. Following the terminology established in previous work, we refer to galaxies formed through monolithic collapse as "model relics" \citep{2022MNRAS.516.1081E, 2024MNRAS.528.4264E} and denote galaxies formed via mergers of two such relics as "merged models." This work extends the analysis of \citet{2022MNRAS.516.1081E} by using these model relics to investigate how mergers influence galaxy structure and scaling relations.

The model relics originate from hydrodynamical simulations in the MOND framework, where non-rotating post-Big-Bang gas clouds collapse in isolation without considering cosmic expansion. These models evolve for 10 Gyr, producing galaxies with star formation histories comparable to observed early-type galaxies (ETGs). However, assuming the nominal age of the Universe to be the standard 13.8 Gyr \citep{2020A&A...641A...6P} their formation times are slightly later than those inferred from observations, suggesting additional environmental factors may have accelerated early star formation in the real Universe, or an older Universe. Further details of the simulation setup can be found in \citet{2022MNRAS.516.1081E}.

Here, we conduct additional simulations to explore the impact of galaxy mergers on structural properties. Each simulation begins with a pair of model relics, initially separated by 100 kpc within a 1 Mpc simulation box, such that they collide and merge within $\approx$ 1 Myr. The orientations of the relics are randomized to ensure a diverse set of merger scenarios. The goal is to explore the range of possible outcomes without imposing a predetermined interaction geometry. The model relics used in this study (labeled e35, e36, e37, e38, and e39) are drawn from \citet{2022MNRAS.516.1081E} and follow the observed downsizing relation, where more massive galaxies form earlier and more rapidly than lower-mass galaxies (see \citealt{2010MNRAS.404.1775T}). The resulting merged models (M1–M6) are formed by merging various combinations of relics, as outlined in Table \ref{tab:two}. Additionally, specific relics (Re36, Re38, and Re39) are rotated by 45 degrees relative to the XY plane before the merger.

The model relics are stellar-only models, formed from the collapse of post-Big-Bang gas clouds using the PoR code (see Section 2) within a MOND framework, where star formation is triggered when the gas density exceeds a threshold \citep{2015CaJPh..93..232L, 2021arXiv210111011N}. Readers interested in the full simulation methodology may refer to \citet{2022MNRAS.516.1081E} and related works \citep{2015CaJPh..93..232L, 2020ApJ...890..173W, 2021arXiv210111011N}.

\subsection{Luminosities and surface brightness}

We begin by determining the center of the galaxy by computing the mass-weighted average positions of particles within the region of interest. The coordinates of the galaxy center, $x_{\text{center}}$ and $y_{\text{center}}$, are given by
\begin{equation}
    x_{\text{center}} = \frac{\sum_i m_i x_i}{\sum_i m_i}, \quad y_{\text{center}} = \frac{\sum_i m_i y_i}{\sum_i m_i},
\end{equation}
where $m_i$ is the mass of the $i$-th particle, and $x_i$ and $y_i$ are its projected coordinates on the $xy$-plane. Once the galaxy center is established, we calculate the projected radial distance $r_i$ of each particle from the center as
\begin{equation}
    r_i = \sqrt{(x_i - x_{\text{center}})^2 + (y_i - y_{\text{center}})^2}.
\end{equation}

We use the following equations and conversions required to conduct the comparison study. The luminosity in the K-band (\( L_{\rm K} \)) is calculated from the final stellar mass of the galaxy (\( M_{\rm final} \)) using a constant mass-to-light ratio (\( \gamma \)), assumed to be 0.8 for the K-band:

\begin{equation}
\label{eq:luminosity}
    L_{\rm K} = \frac{M_{\rm final}}{\gamma}.
\end{equation}
This value is consistent with observationally inferred values for early-type galaxies in the near-infrared (e.g., \citealt{2001ApJ...550..212B, 2003ApJS..149..289B, 2006MNRAS.366.1126C}). While the mass-to-light ratio can vary depending on the stellar population, metallicity, and star formation history, a value of \(\gamma \approx 0.8\) is typically found for quiescent, massive elliptical galaxies, particularly in the MOND framework where dark matter is not included in mass estimates. Given that our model galaxies form from a monolithic collapse and produce compact, massive relics comparable to observed early-type galaxies, this assumption provides a reasonable first-order approximation. However, variations in the mass-to-light ratio due to age, metallicity and galaxy-wide stellar initial mass function gradients \citep{2024arXiv241007311K} may introduce minor systematic effects, which should be considered in future studies.

The surface brightness is calculated as:

\begin{equation}
\label{eq:SB1}
    I_{\rm eff} = \frac{L_{\rm K}}{2 \pi R_{\rm eff}^{2}},
\end{equation}
where \( R_{\rm eff} \) is the projected effective radius, defined as the radius within which half of the final stellar mass of the galaxy is enclosed. The surface brightness is then converted to magnitudes per square arcsecond using:
\begin{equation}
\label{eq:SB2}
    \langle\mu_{\rm K}\rangle_{\rm eff} (\mathrm{ mag \ arcsec^{-2}}) = M_{\rm \odot K} + 21.572 - 2.5 \log_{10} (I_{\rm eff}/\rm L_{\odot} \rm pc^{-2}),
\end{equation}
where \( M_{\rm \odot K} \) is the absolute magnitude of the Sun in the K-band (\( = 3.28 \)), a near-infrared band commonly used for studying the stellar populations and mass distributions of galaxies \citep{2010gfe..book.....M}. The absolute magnitude of the galaxy is given by:
\begin{equation}
\label{eq:absolutemagnitude}
    M_{\rm K} = -2.5 \log_{10} (L_{\rm K}/L_{\rm \odot}) + M_{\rm \odot K}.
\end{equation}

\begin{table}
	\begin{adjustwidth}{-\extralength}{0cm}

\centering
	\caption{Properties of relics and mergers (for edge-on projections). The relics e35 - e39 are published by \citet{2022MNRAS.516.1081E}.}
	\label{tab:two}
\begin{tabular}{c c c c c c c c c c}

\hline
Model & $M_{\rm final}$ & $V_{\sigma}$ & $R_{\rm eff}$ & ellipticity & $\mu_{\rm eff}$ & $M_{\rm K}$ & $\rm log_{10}$ ($f_{\rm eff}$) & $\lambda_{R}$ & $(A_{4}/A)$ \\
Name & (10$^{11}$ $M_\odot$) & (km/s) & (kpc) & $\epsilon$ & ($\rm mag arcsec^{-2}$) & (mag) & & & \\
\hline
\\
M1(e35+e35) & 0.17 & 71.55 & 2.65 & 0.27 & 18.14 & -22.53 & -2.8 & 2.43 & 0.07 \\
M2(e36+e36) & 0.38 & 103.28 & 2.94 & 0.41 & 17.49 & -23.41 & -3 & 2.37 & 0.02 \\
M3(e37+e37) & 0.9 & 139.63 & 3.25 & 0.27 & 16.77 & -24.34 & -3.23 & 2.26& 0.12 \\
M4(e38+e38) & 1.25 & 153.73 & 3.05 & 0.25 & 16.28 & -24.7 & -3.34 & 2.41 & -0.02 \\
M5(Re36+Re39) & 0.75 & 163.4 & 2.49 & 0.22 & 16.39 & -24.14 & -3 & 1.66 & -0.06 \\
M6(Re38+Re38) & 0.98 & 178.6 & 3.61 & 0.17 & 16.91 & -24.44 & -3.37 & 2.94 & -0.01 \\
e35 & 0.1 & 112.73 & 0.47 & 0.73 & 14.93 & -21.96 & -1.38 & 0.27& 2.40 \\
e36 & 0.3 & 190.18 & 0.6 & 0.74 & 14.33 & -23.15 & -1.85 & 0.35& 1.19 \\
e37 & 0.5 & 234.48 & 0.84 & 0.68 & 14.47 & -23.71 & -2.2 & 0.46 & 4.16 \\
e38 & 0.64 & 275.06 & 0.95 & 0.65 & 14.62 & -23.97 & -2.38 & 0.56 & 2.13 \\
e39 & 0.7 & 369.5 & 1.04 & 0.74 & 14.38 & -24.07 & -2.52 & 0.52 & 1.80 \\
\hline
\\

\end{tabular}
\\Note: Column 1: name -- M1, M2, M3, M4, M5 and M6 are the merged models. The brackets show the relics that are used to form the merged model and Re36, Re38 and Re39 indicate the relics that were placed in the simulation box rotated at different angles (where parent models are placed tilted at an angle of 45 deg with respect to the XY plane before the merger event), Column 2 ($M_{\rm final}$): final mass of the galaxy, Column 3: velocity dispersion at effective-radius, Column 4 ($R_{\rm eff}$): projected effective-radius, Column 5 ($\epsilon$): the projected ellipticity, Column 6 ($\mu_{\rm eff}$): surface brightness at effective-radius, Column 7 ($M_{\rm K}$): absolute magnitude, Column 8 ($f_{\rm eff}$): effetive-phase-space density, Column 9 ($\lambda_{R}$): rotation parameter and Column 10 ($A_{4}/A$): isophotal parameter.
	\end{adjustwidth}

\end{table}

\subsection{Shape in formation}

To determine the elliptical shape, or ellipticity of the galaxy’s projection, we first calculate the second moments of the mass distribution in the projected plane. For each stellar particle, the moments are computed using their positions \((x, y)\) in the x-y sky plane. The moment matrix elements are calculated as,

\begin{equation}
\begin{aligned}
M_{xx} &= \frac{\sum_{j=1}^{N} m_j x_j^2}{\sum_{j=1}^{N} m_j}, \\
M_{yy} &= \frac{\sum_{j=1}^{N} m_j y_j^2}{\sum_{j=1}^{N} m_j}, \\
M_{xy} &= \frac{\sum_{j=1}^{N} m_j x_j y_j}{\sum_{j=1}^{N} m_j}
\end{aligned}
\end{equation}
where $m_{\rm j}$ is the mass of stellar particle $j$ and $N$ is the total number of stellar particles. Using these elements, the semi-major (\(a\)) and semi-minor (\(b\)) axes of the projected ellipse are derived from the eigenvalues of the moment matrix, given by

\begin{equation}
\begin{aligned}
a^2 = \frac{1}{2} \left( M_{xx} + M_{yy} + \sqrt{(M_{xx} - M_{yy})^2 + 4M_{xy}^2} \right)\\
b^2 = \frac{1}{2} \left( M_{xx} + M_{yy} - \sqrt{(M_{xx} - M_{yy})^2 + 4M_{xy}^2} \right).
\end{aligned}
\end{equation}
This follows from the moment of inertia tensor formalism, where the eigenvalues of the second moment matrix define the principal axes of the projected mass distribution (see \citealt{2008gady.book.....B}). The ellipticity \(\epsilon\) is then calculated as \(\epsilon = 1 - \frac{b}{a}\). 

These quantities characterize the shape and orientation of the galaxy in the projected plane. To analyze the kinematic properties of a galaxy, the stellar population is divided into concentric annular regions centered on the galactic core by calculating the projected radial distance of each star from the galactic center and defining the boundaries for the annuli, such that the edges are established to create regions of equal radial width. For each annulus, stellar particles are identified that fall within its radial boundaries. The line-of-sight velocity dispersion and rotation velocity are then calculated for each annular region. The line-of-sight velocity dispersion (z-direction), which quantifies the spread of velocities along the line of sight, is computed as
\begin{equation}
V_{\sigma} = \sqrt{\frac{1}{N-1} \sum_{j=1}^{N} (v_{z,j} - \overline{v_z})^2},
\end{equation}
where \(N\) is the number of stellar particles in the annulus, and \(\overline{v_z}\) is the mean LOS velocity within the annulus. The rotation velocity is determined from the mean tangential component of the stellar particle velocities in an annulus. The tangential velocity of an individual particle $j$ is computed as:
\begin{equation}
v_{\text{tan},j} = \frac{x_j v_{y,j} - y_j v_{x,j}}{r_j}.
\end{equation}
The rotation velocity for the annulus is then given by the mean tangential velocity,
\begin{equation}
V_{\text{rot}} = \frac{1}{N} \sum_{j=1}^{N} v_{\text{tan},j}.
\end{equation}

\begin{figure}
\includegraphics[width=10.5 cm]{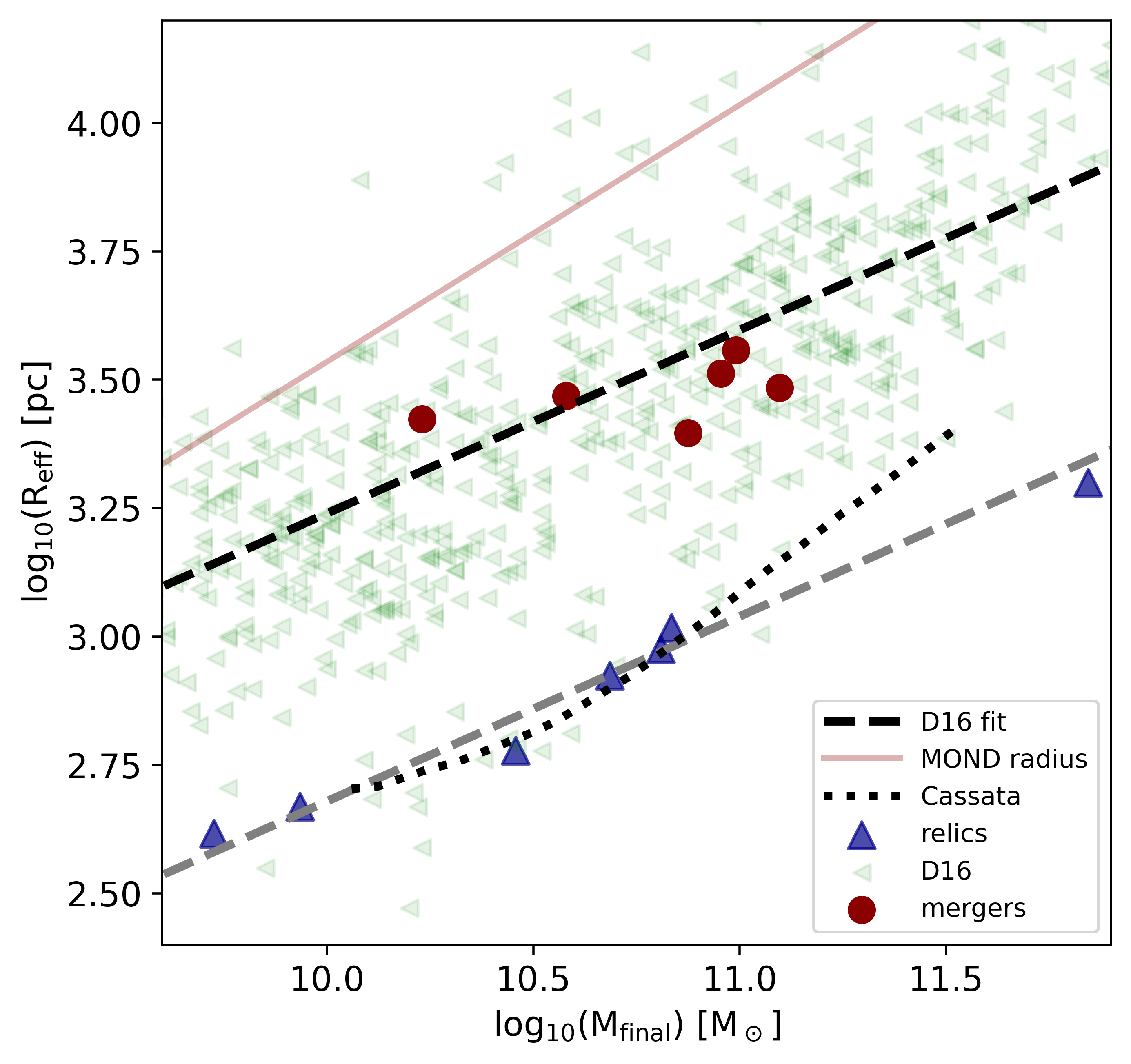}
    \caption{The effective radius -- mass relation from \citet{2022MNRAS.516.1081E}. The green data points are the observed ETGs from \citealt{2016MNRAS.460.4492D}(D16) and the dashed black line is the best-fit relation for the ETGs at z=0. The red circles are the mergers computed here and blue triangles are the relics from \citet{2022MNRAS.516.1081E} and the grey dashed line is the best-fit for the relics. We over-plotted the best-fit relation for galaxies at 1.2 $<$ z $<$ 2.5 from \citet{2011ApJ...743...96C}.} The MOND radius, \( r_{\rm {M}} = \left(\frac{\mathrm{G} \cdot M_{\rm final}}{a_{0}}\right)^{0.5} \) (equation 13 in \citealt{2022MNRAS.516.1081E}), is also shown.

    \label{fig:reffdab}
\end{figure} 

We reproduce Fig. 7 from \citet{2022MNRAS.516.1081E} in Figure \ref{fig:reffdab} to emphasize the mass-size relation of the galaxies formed in our simulations, alongside observational data. \citet{2011ApJ...743...96C} notes that ETGs at z > 1 are generally smaller, and their size increases as redshift decreases. We overlay the mass-size relation for galaxies at 1.2 $<$ z $<$ 2.5 from \citet{2011ApJ...743...96C} and find a comparable relation to the relics in our study. This suggests that the mass-size relation is a reasonable approximation within the MOND framework. Consequently, ETG formation in MOND likely involves a two-phase process: an initial monolithic collapse followed by a subsequent phase of few dry mergers, which further increases the galaxy size. This is a likely occurence because elliptical galaxies are found at the centres of galaxy clusters where most likely multiple such galaxies formed with some of them merging later given the high galaxy number densities.

\section{Results}
\label{sec:results}

Here we compare the results of the mergers and the relics with the observed relations on ETGs. The data for the observed ETGs are taken from \citet{1983ApJ...266...41D}, \citet{2001MNRAS.326..473H}, \citet{2006AJ....131..185R} and \citet{2007MNRAS.379..401E}.

\subsection{Ellipticity relation}
\label{sec:4.1}

\begin{figure}
\begin{adjustwidth}{-\extralength}{0cm}
\centering
\includegraphics[scale=0.6]{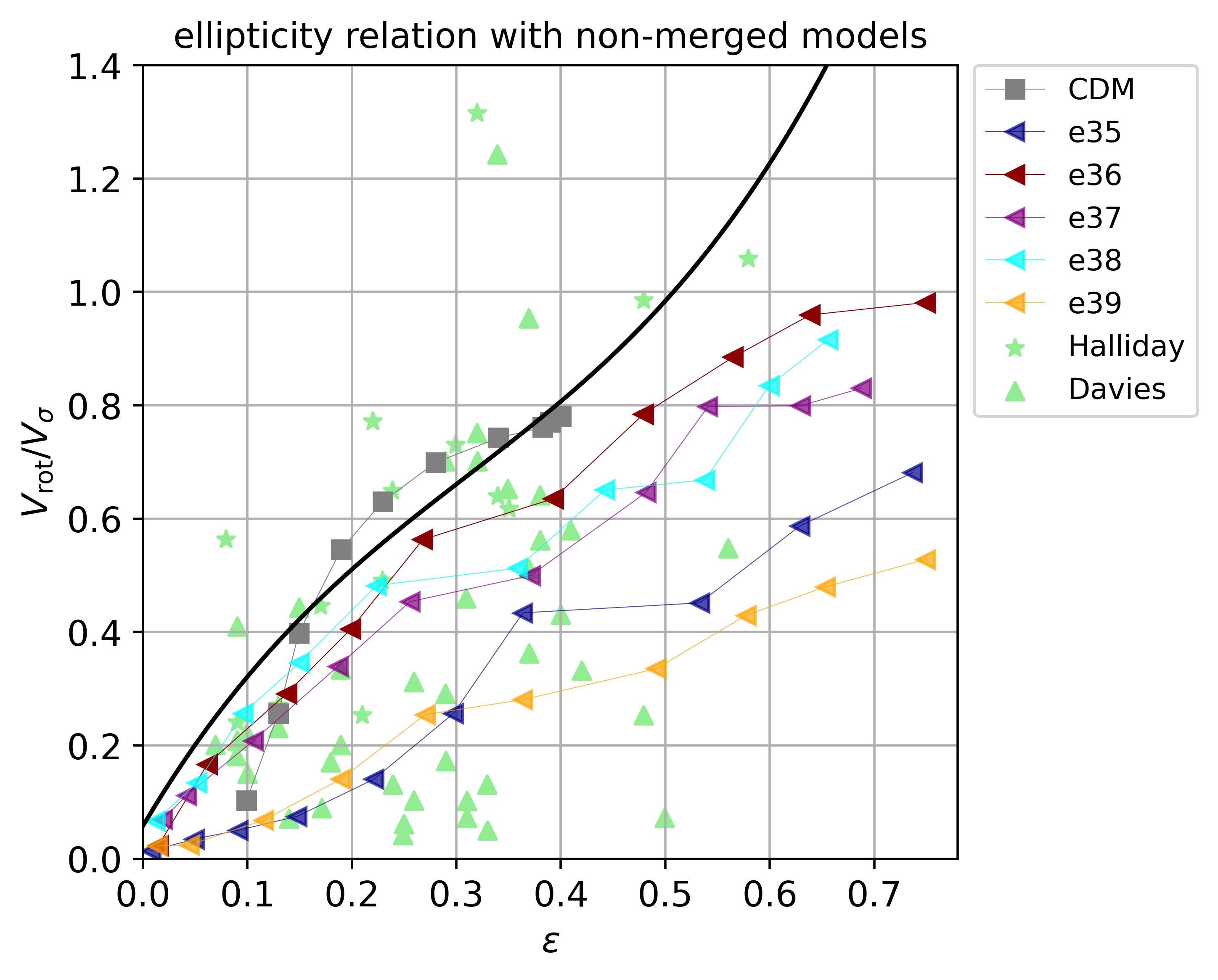}
\includegraphics[scale=0.6]{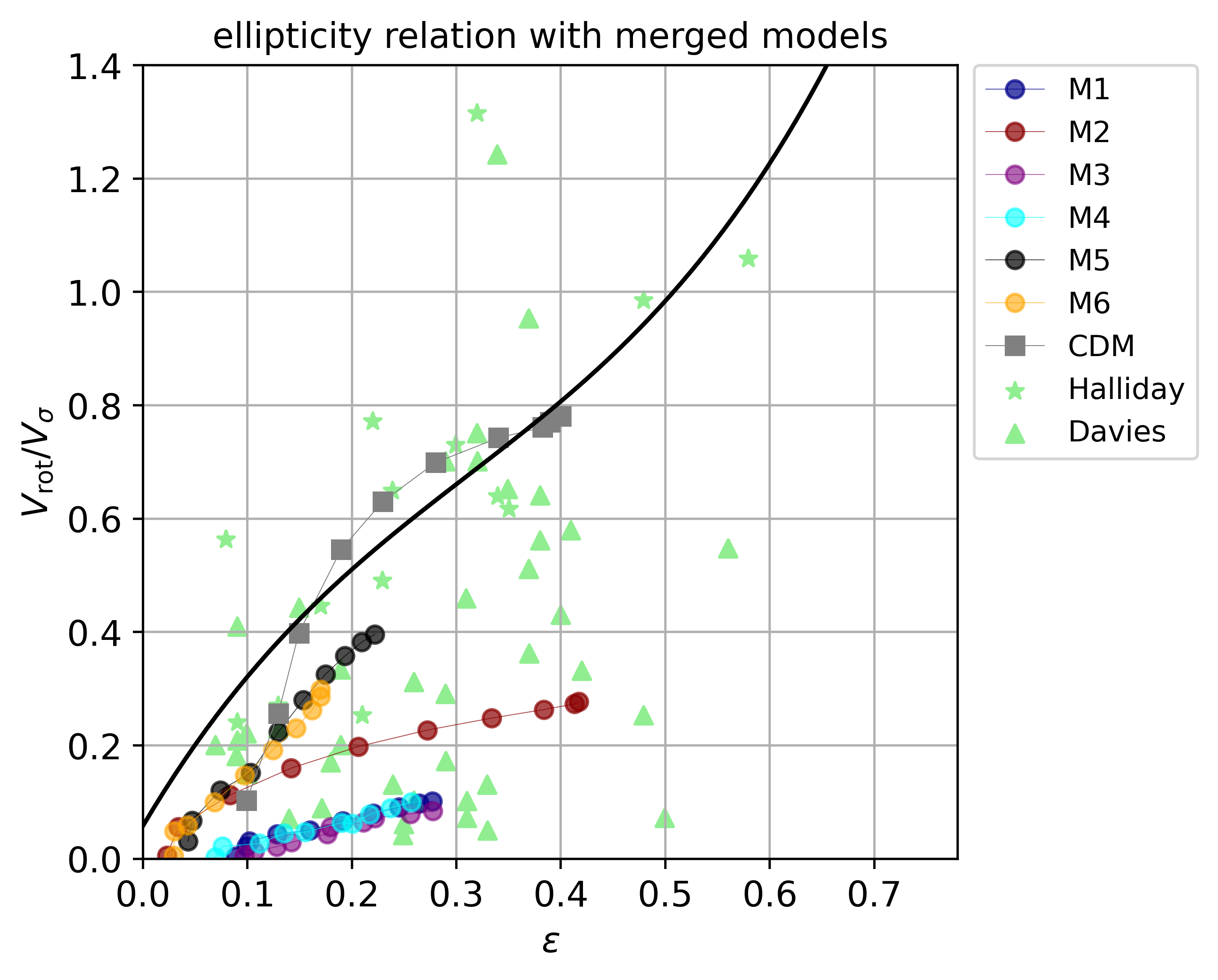}
\\
\end{adjustwidth}
    \caption{The $V_{\rm rot}/V_{\sigma}$ -- ellipticity relation plotted at effective radius for the model relics (left panel, e35-e39) and merged models (right panel, M1 - M6) at different projections. The properties of the merged models are provided in Table \ref{tab:two}. Similarly, an ETG formed in a $\Lambda$CDM simulation (\citealt{2003ApJ...590..619M}) is plotted in gray squares, each square being a particular projection. The solid black line indicates the relation expected for an oblate rotator whose shape is determined by the degree of rotational support (see Figs. 4–6 of \citealt{1987gady.book.....B}). The filled green coloured stars are the ellipticals from \citet{2001MNRAS.326..473H}. The filled green triangles are for spheroidal systems (including elliptical galaxies and the bulges of disk galaxies) of low luminosity and giant ellipticals respectively from \citet{1983ApJ...266...41D}.} 

    \label{fig:ellipticityrelation}
\end{figure}  

Figure \ref{fig:ellipticityrelation} illustrates the $V_{\rm rot}/V_{\sigma}$ - ellipticity relation ($V_{\rm rot}/V_{\sigma}$ vs. $\epsilon$). The green triangles and stars represent observed ETGs, respectively, from \citet{1983ApJ...266...41D} and \citet{2001MNRAS.326..473H}. The colored circles indicate mergers viewed from different projections. Similarly, left-pointing triangles represent the relics at various projections. For comparison, gray squares denote projections for a galaxy formed in a $\Lambda$CDM hydrodynamical simulation in different projection \citep{2003ApJ...590..619M}.

The mergers exhibit an ellipticity extending up to 0.4, whereas the relics demonstrate a higher ellipticity, reaching up to 0.79. This difference is due to the compact and elongated nature of the relics, which contributes to their greater ellipticity \citep{2024MNRAS.528.4264E}. Additionally, $V_{\rm rot}/V_{\sigma}$ for the merged models varies between 0 and 0.4. This range indicates a significant degree of randomness in the motion of the stellar particles within the merged galaxies experiencing a redistribution of angular momentum and kinetic energy during the merging process. This redistribution can lead to more isotropic or randomized stellar orbits, thereby reducing the overall ellipticity. In contrast, higher $V_{\rm rot}/V_{\sigma}$ values would indicate more coherent rotational motion, which is observed for the relics. The results suggest that the elliptical characteristics of ETGs can be achieved through the merger of two galaxies, each formed via the monolithic collapse of a post-Big-Bang gas cloud with negligible initial rotation.

\subsection{Rotation parameter $\lambda_{R}$}
Building on the $V_{\rm rot}/V_{\sigma}$ versus ellipticity results presented above, we further quantify the galaxy’s rotational support using the $\lambda_{R}$ parameter, as described in \citet{2007MNRAS.379..401E}. This metric has gained widespread use in observational studies as a robust measure of rotational support, particularly given its applicability when tangential velocity components are unavailable. Including $\lambda_{R}$ in our analysis ensures that our findings are directly comparable with contemporary datasets, providing a consistent framework that aligns with current observational methodologies.

The $\lambda_{R}$ parameter quantifies the degree of ordered rotation relative to random motion in a galaxy. Following the formalism of \citet{2007MNRAS.379..401E}, it is defined as

\begin{equation}
    \lambda_{R} = \sum_{i}^{N}\frac{m_i R_i |V_{{\rm los},i}|}{m_i R_i \sqrt{V_{{\rm los},i}^2 + V_{\sigma,i}^2}},
\end{equation}
where $m_i$ is the mass of the $i$th stellar particle, $R_i$ is its projected radius in the XY plane, $V_{{\rm los},i}$ is the line-of-sight velocity, and $V_{\sigma,i}$ represents the local velocity dispersion within radial bins centered at $R_i$. This formulation accounts for both rotationally and dispersion-supported contributions, enabling a distinction between slow and fast rotators.

\begin{figure}
\includegraphics[width=10.5 cm]{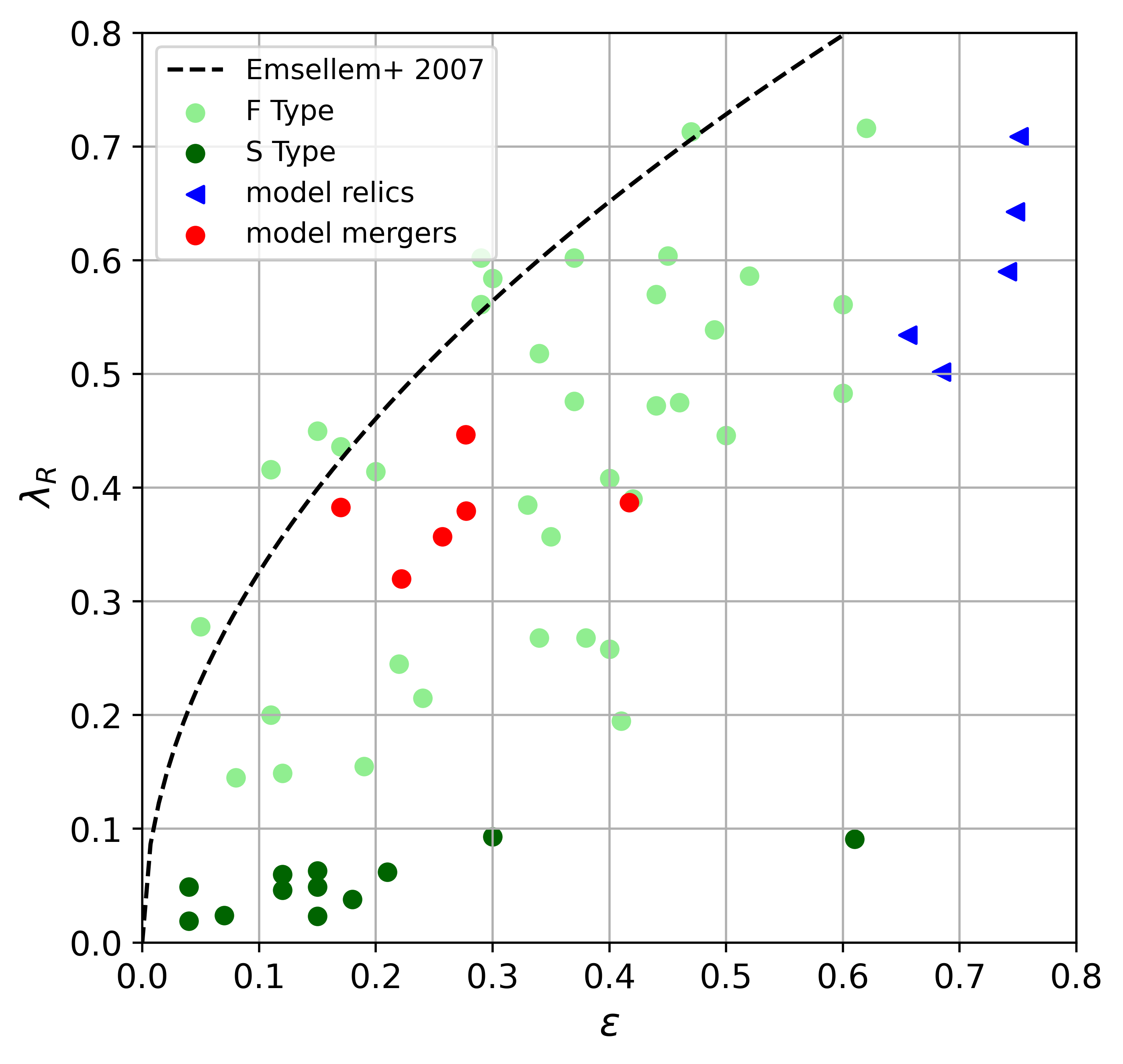}
    \caption{The rotational support parameter, $\lambda_{R}$, and ellipticity, $\epsilon$, are presented for both simulated and observed galaxies from \citet{2007MNRAS.379..401E}. Observational data is represented by green points, with light green indicating fast rotators (F Type) and dark green indicating slow rotators (S Type). The model relics are represented by blue triangles, and model mergers by red markers. The dashed line, derived from \citet{2007MNRAS.379..401E}, represents the theoretical boundary for isotropic oblate rotators.}
    \label{fig:lambdaR1}
\end{figure}

The line-of-sight velocity $V_{{\rm los},i}$ is computed as the projection of the stellar velocity onto the observer's viewing direction. Assuming an observer aligned with the z-axis, this simplifies to

\begin{equation}
    V_{{\rm los},i} = v_{z,i}.
\end{equation}
Figure \ref{fig:lambdaR1} presents the relationship between the rotational support parameter $\lambda_{R}$ and ellipticity $\epsilon$ for model relics and merged models in this study, compared to observed galaxies from \citet{2007MNRAS.379..401E}. The observational data is represented by green markers, with light green indicating fast rotators (F-type) and dark green indicating slow rotators (S-type). The model relics are plotted as blue triangles, while the model mergers are represented by red points. The dashed line, derived from \citet{2007MNRAS.379..401E}, represents the theoretical boundary for isotropic oblate rotators.

The merged models (red points) exhibit a range of $\lambda_{R}$ values and ellipticities, primarily occupying an intermediate region between observed slow and fast rotators. Some merged models exhibit moderate $\lambda_{R}$ values and ellipticities comparable to observed fast rotators, suggesting that certain merger events can maintain or induce significant rotational structure. Other merged models, particularly those with lower $\lambda_{R}$ values, align more closely with the properties of observed slow rotators. The model relics (blue triangles) show higher $\lambda_{R}$ values and ellipticities, placing them among the fast rotators in the observational dataset \citep{2007MNRAS.379..401E}. This pattern indicates that model relics are predominantly rotation-supported, with relatively high ellipticities suggestive of a disky morphology.

\subsection{Isophotal analysis}

\begin{figure}
\includegraphics[width=9 cm]{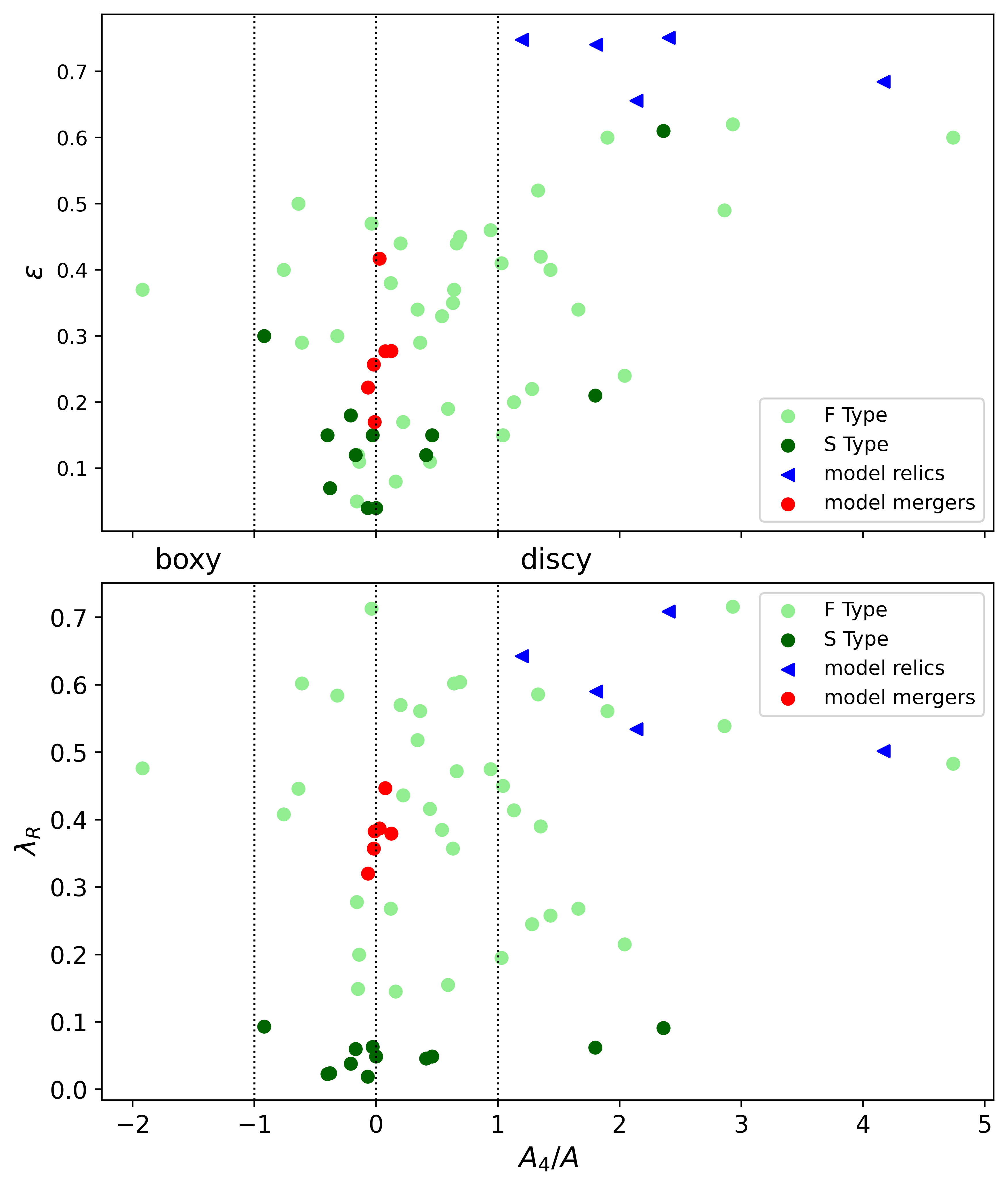}
    \caption{Isophotal shape parameter $(A_4/A)$ versus ellipticity $\epsilon$ (top panel) and $(A_4/A)$ versus rotational support parameter $\lambda_{R}$ (bottom panel). The green data points are from observations \citep{2007MNRAS.379..401E}, while the blue triangles and red circles represent the relics and mergers in this work.}
    \label{fig:lambdaR2}
\end{figure} 

\begin{figure}
\begin{adjustwidth}{-\extralength}{0cm}
\centering
\includegraphics[scale=0.31]{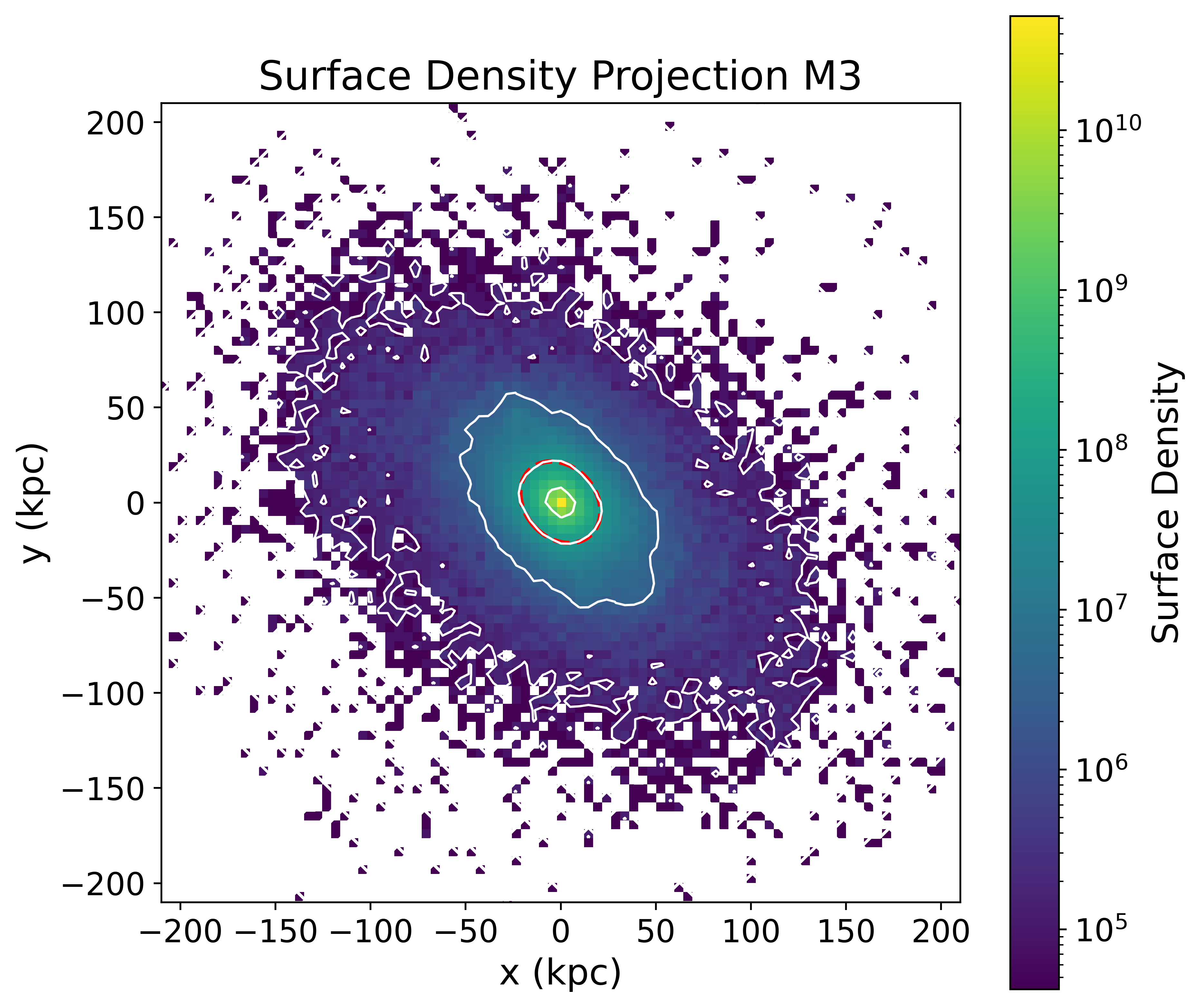}
\includegraphics[scale=0.31]{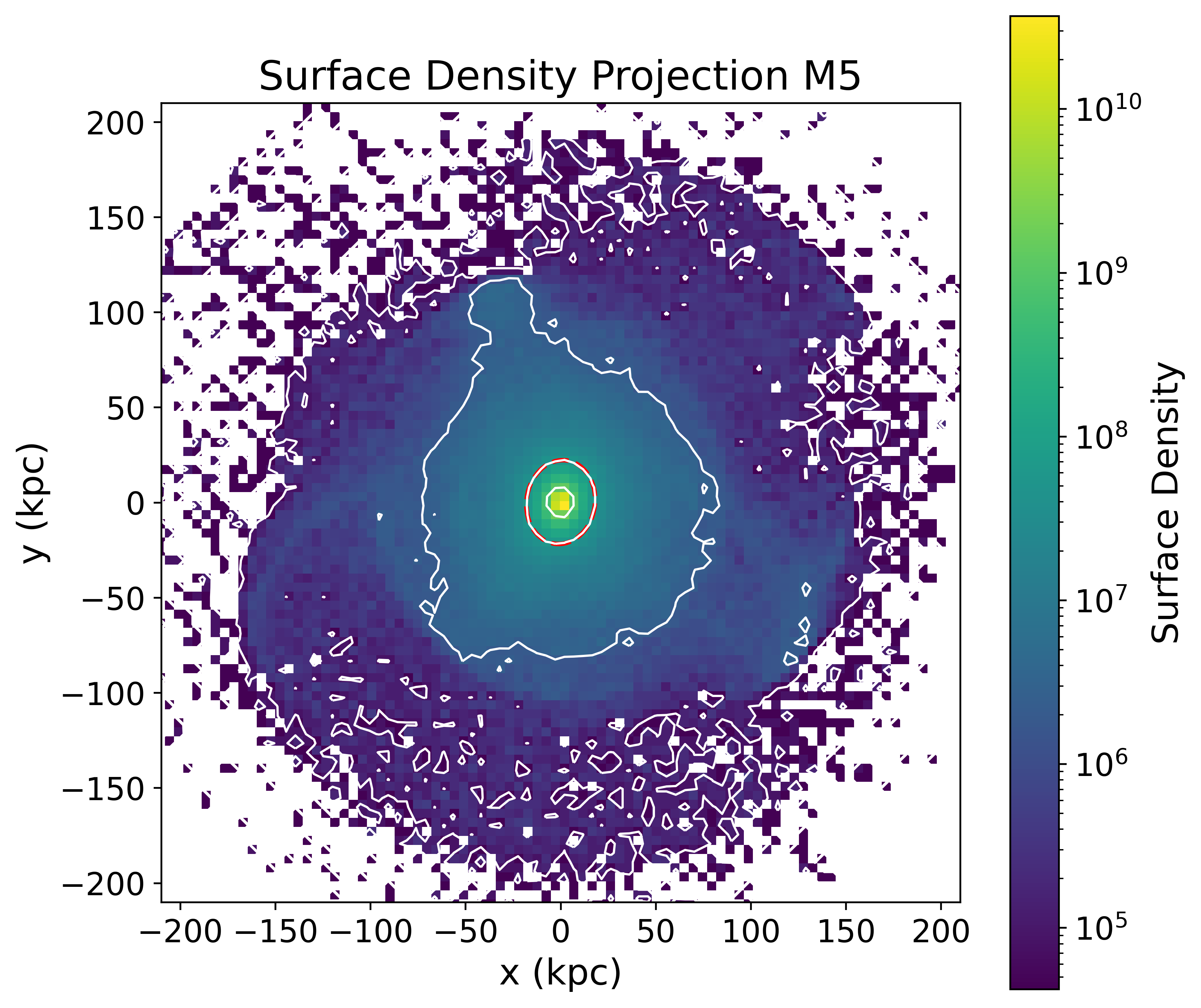}
\includegraphics[scale=0.31]{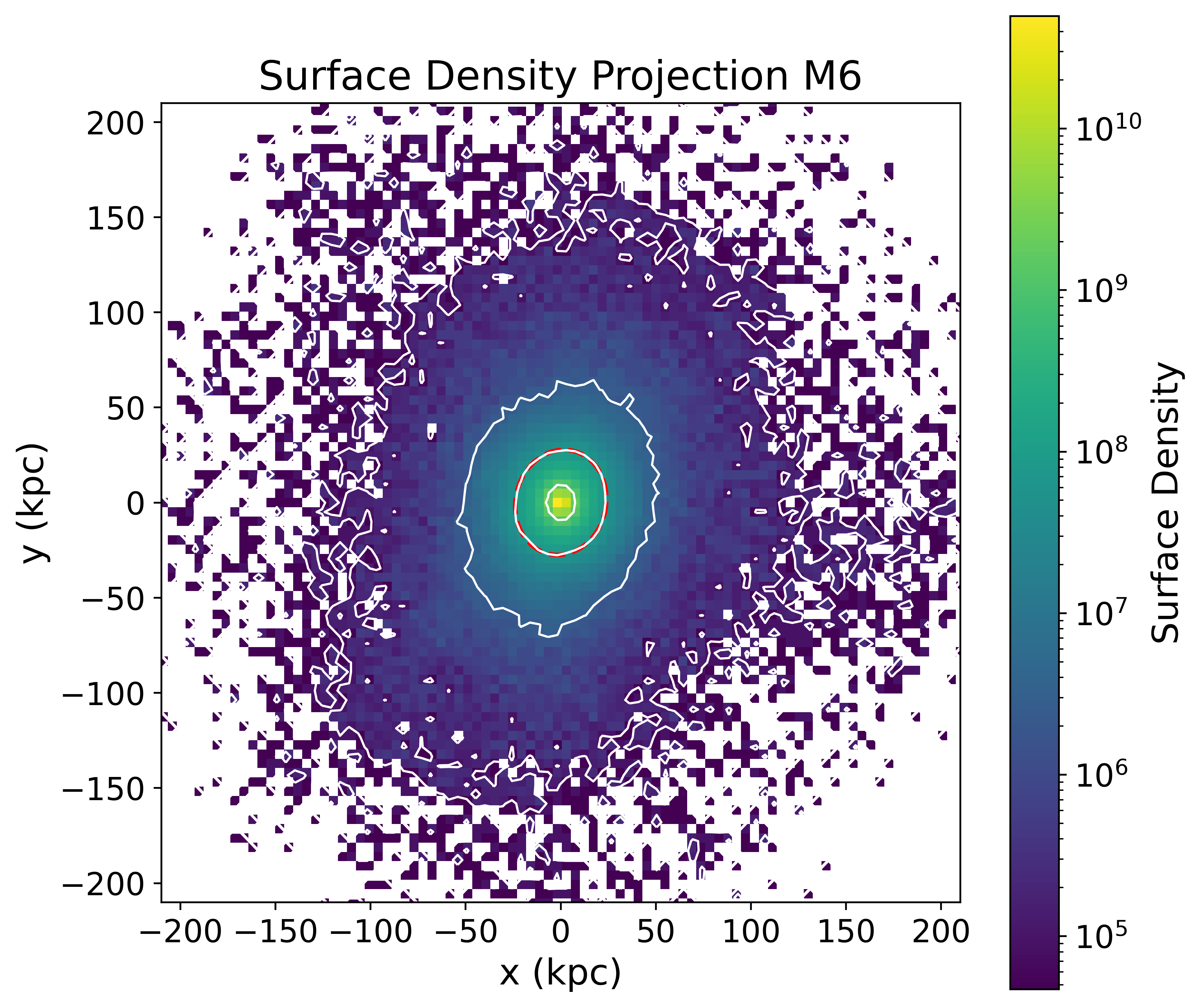}
\\
\includegraphics[scale=0.31]{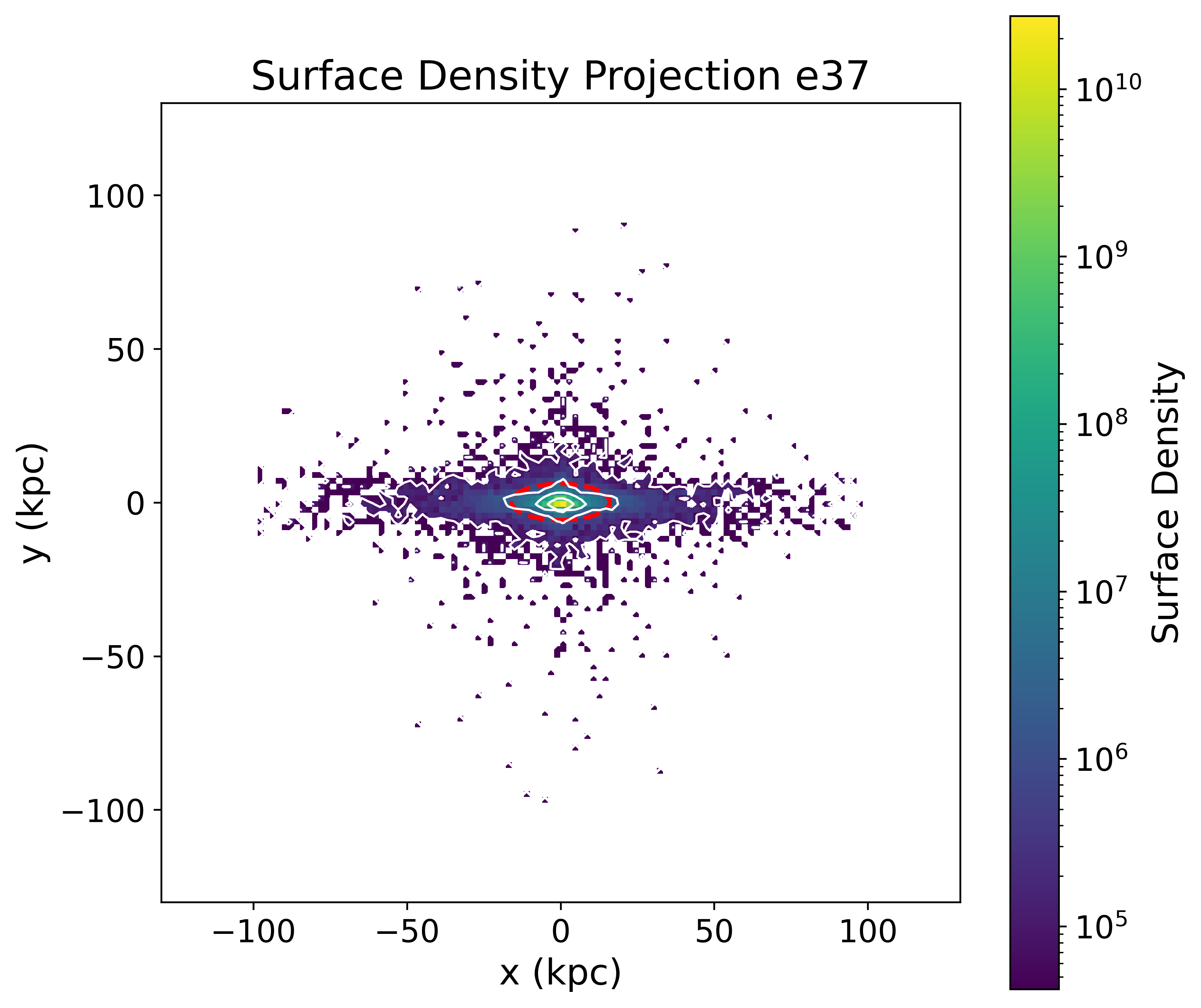}
\includegraphics[scale=0.31]{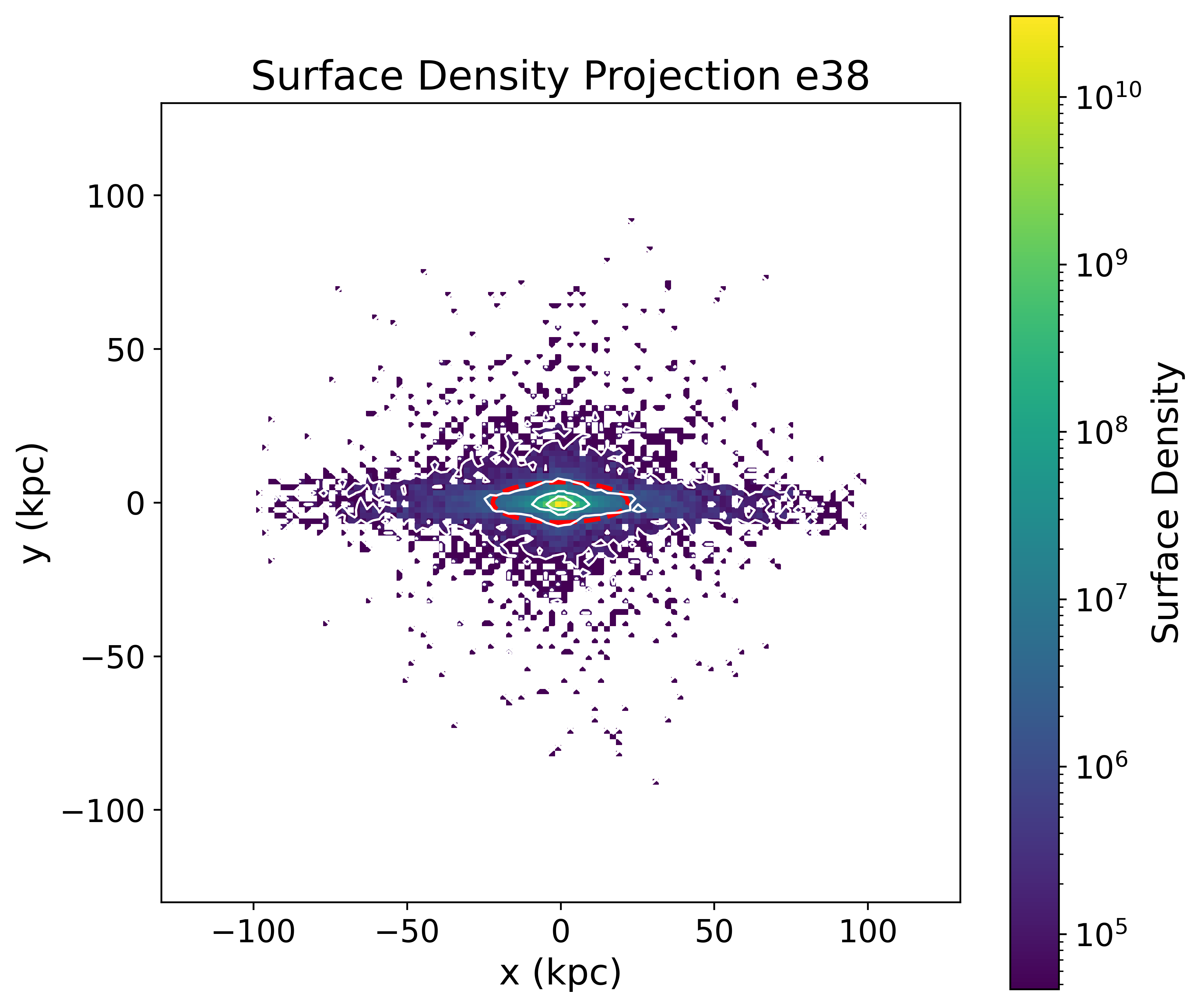}
\includegraphics[scale=0.31]{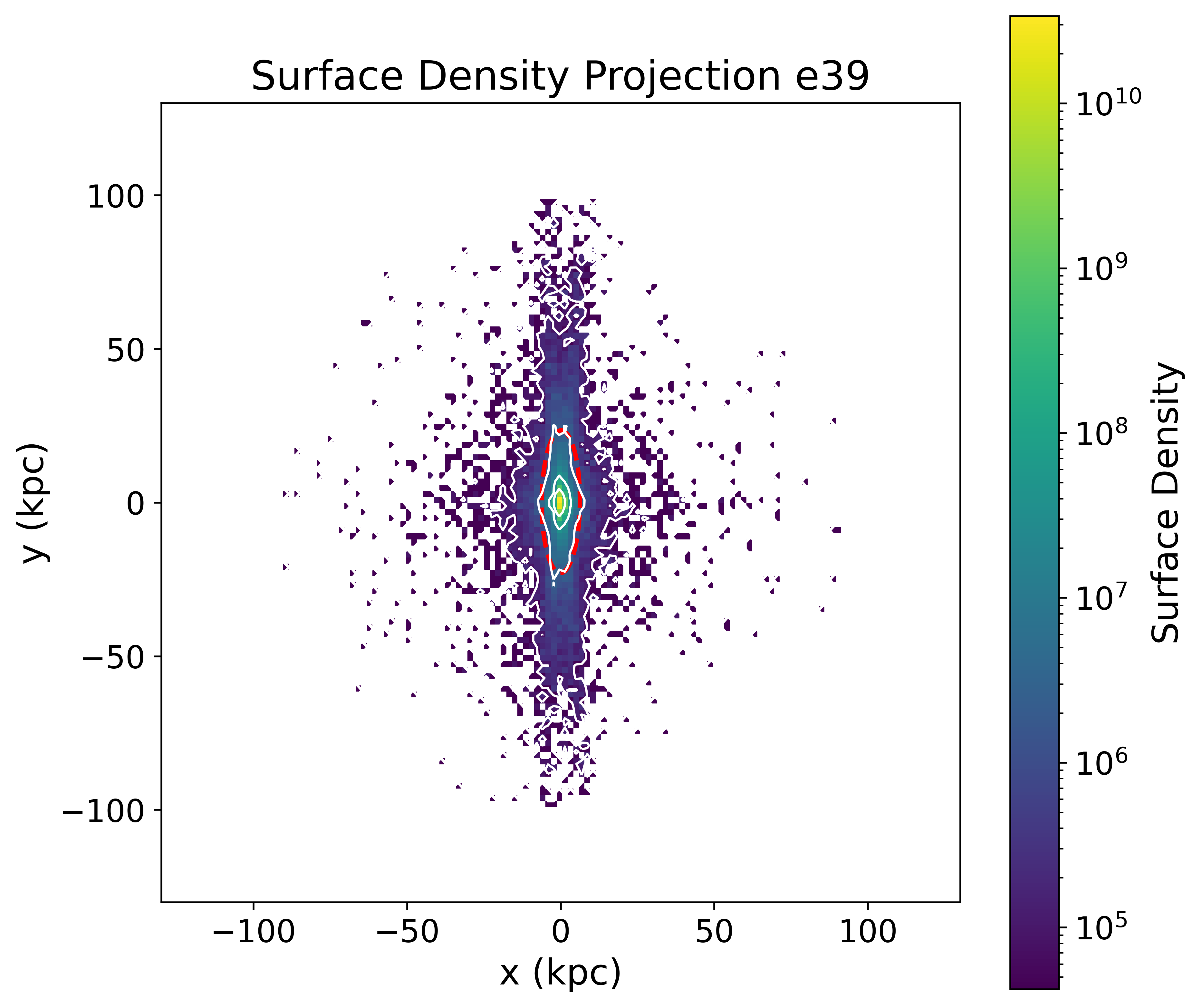}
\\
\end{adjustwidth}
\caption{Surface density plots with isophotal contours of three mergers (M3, M5, and M6 from left to right) in the top panel and three relics (e37, e38, and e39 from left to right) in the bottom panel.}
\label{fig:isophotal}
\end{figure}

The isophotal distortion parameter \( A_4/A \) (in Figure \ref{fig:lambdaR2}, and formally defined below) quantifies deviations of isophotal shapes from perfect ellipses. To compute this parameter, we project the stellar distribution, generate isophotes, fit ellipses to the isophotes (Figure \ref{fig:isophotal}), and calculate radial deviations using Fourier analysis \citep{1988A&AS...74..385B}. A 2D histogram of the stellar particle positions \( (x', y') \) is constructed to derive the surface density \( \Sigma(x', y') \). Isophotes are then defined as curves of constant surface density \( \Sigma_i \), allowing for an analysis of the projected shape.

To quantify deviations from a perfect elliptical shape, we fit ellipses to the isophotes. The largest path of each isophote is extracted, and an ellipse is fitted using the general equation:

\begin{equation}
\frac{((x - h) \cos(\theta) + (y - k) \sin(\theta))^2}{A^2} + \frac{((x - h) \sin(\theta) - (y - k) \cos(\theta))^2}{B^2} = 1,
\end{equation}

where \( A \) and \( B \) are the semi-major and semi-minor axes of the fitted ellipse, respectively, \( (h, k) \) is the center of the ellipse, and \( \theta \) is the orientation angle. Here, \( x \) and \( y \) denote the Cartesian coordinates of each isophote point in the projected plane. The optimal parameters \( A \), \( B \), \( h \), \( k \), and \( \theta \) are determined via non-linear least squares optimization.


After fitting, we quantify the radial deviation of the isophotal shape from the fitted ellipse. For each isophote point \( (x_i, y_i) \), the projected radial distance \( R_i \) from the center \( (h, k) \) is computed as:
\begin{equation}
R_i = \sqrt{(x_i - h)^2 + (y_i - k)^2}.
\end{equation}
The mean radial distance \( \bar{R} \) is then defined as:
\begin{equation}
\bar{R} = \frac{1}{N} \sum_{i=1}^N R_i,
\end{equation}
where \( N \) is the number of points in the isophote path. The deviation \( \delta R_i \) of each point from the mean radius is given by:
\begin{equation}
\delta R_i = R_i - \bar{R}.
\end{equation}
To extract isophotal shape deviations, a Fourier decomposition of \( \delta R_i \) as a function of the polar angle \( \theta \) is performed:
\begin{equation}
\delta R(\theta) = \sum_{n=0}^\infty A_n \cos(n \theta) + B_n \sin(n \theta).
\end{equation}
The Fourier coefficients are computed as:
\begin{equation}
A_n = \frac{2}{N} \sum_{i=1}^N \delta R_i \cos(n \theta_i), 
\quad 
B_n = \frac{2}{N} \sum_{i=1}^N \delta R_i \sin(n \theta_i)
\end{equation}
where \( \theta_i \) is the polar angle corresponding to the \( i \)-th point on the isophote.
The \( n = 4 \) Fourier mode coefficient, \( A_4 \), quantifies the deviation of the isophote from an ellipse. The isophotal shape parameter (\( A_4/A\)), measures deviations from a purely elliptical shape: positive values (\( A_4/A > 0 \)) correspond to "discy" isophotes, which exhibit an excess of tangential mass distribution, while negative values (\( A_4/A < 0 \)) correspond to "boxy" isophotes, which have more rectangular contours, often associated with mergers or anisotropic stellar orbits. In our study the (\( A_4/A < 0 \)) is calculated at a radial distance equal to the effective radius of the model.

By analyzing multiple isophotes at different surface density levels, this approach provides a detailed characterization of a galaxy's projected shape.

In Figure \ref{fig:lambdaR2}, we explore the relationships between \( (A_4/A) \) and ellipticity \( \epsilon \) (top panel), and between \( (A_4/A) \) and the rotational support parameter \( \lambda_{R} \) (bottom panel). Observational data from \citet{2007MNRAS.379..401E} are plotted in green, with light green indicating fast rotators (F Type) and dark green indicating slow rotators (S Type). Simulated galaxies in this work are represented by red circles (mergers) and blue triangles (relics). The vertical dashed lines at \( a_4/A = \pm 1 \) separate the disky (\( A_4/A > 0 \)) and boxy (\( A_4/A < 0 \)) regions. Merged models (red circles) cluster near \( A_4/A = 0 \), suggesting an overall elliptical morphology. In contrast, model relics (blue triangles) exhibit predominantly positive \( A_4/A \) values, indicating disky structures with significant rotational support.

The bottom panel of Figure \ref{fig:lambdaR2} illustrates the relationship between \( A_4/A \) and \( \lambda_{R} \). The model relics exhibit high \( \lambda_{R} \) values, supporting their classification as rotationally supported disk-like systems. Mergers, however, show a broader range of \( \lambda_{R} \), with most clustering around \( A_4/A \approx 0 \), indicating intermediate isophotal shapes. The observed galaxies with -1 $<$ $A_{4}/A$ $<$ 0.5 and $\lambda R$ $<$ 0.1 indicate them to have formed from more than one merger. Further modelling which is beyond the scope of this study due to computer limitations will test this viable hypothesis.

\subsection{Fundamental Plane}
\label{sec:4.2}

\begin{figure}
\includegraphics[width=10.5 cm]{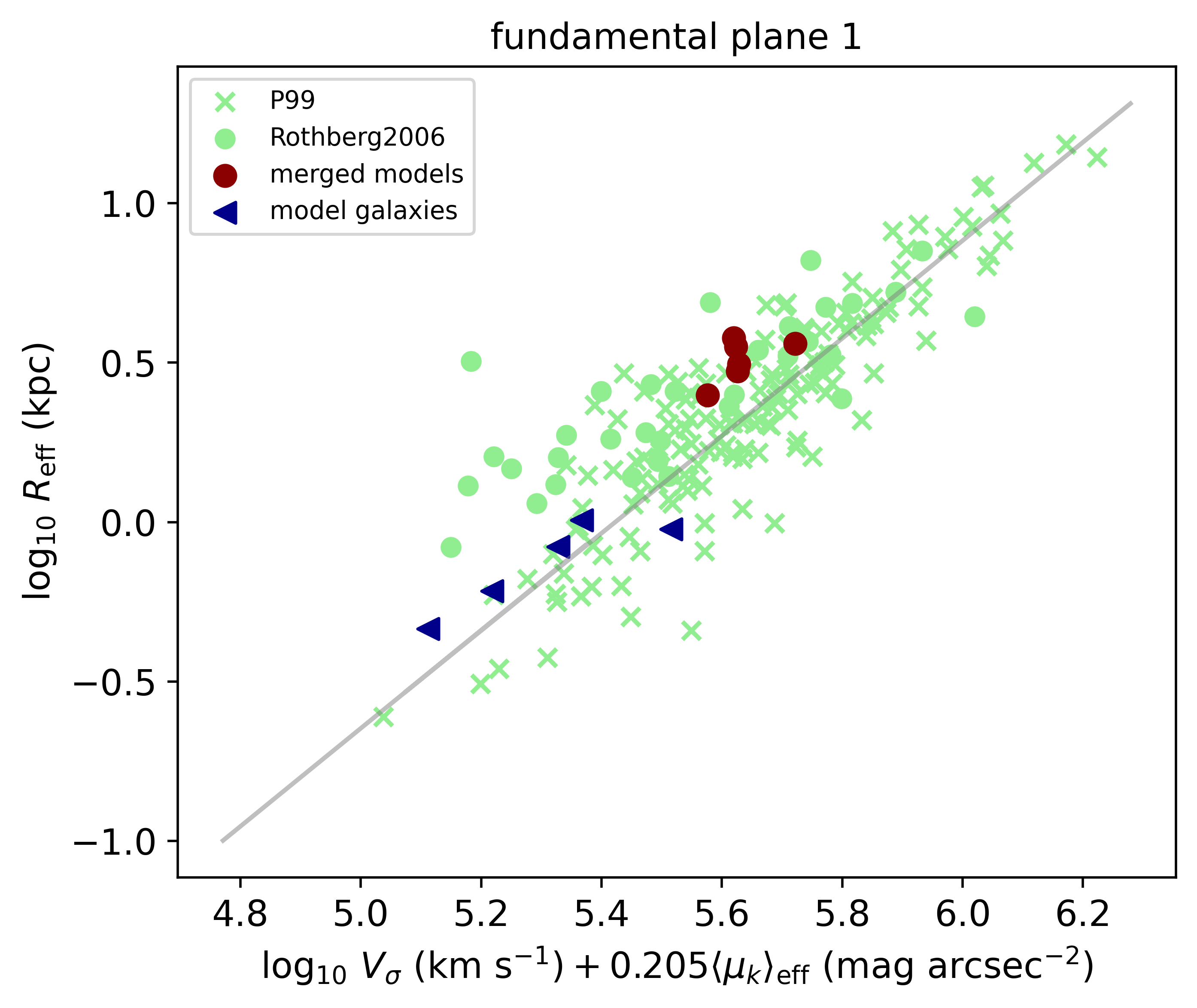}
    \caption{The K-band fundamental plane of elliptical galaxies. The green coloured data plotted are the ellipticals from \citet{1999ApJS..124..127P} and luminous as well as ultra luminous infrared galaxies from \citet{2006AJ....131..185R}. The red filled circles are the mergers and the left pointing blue triangles are the relics (see Table \ref{tab:two}). The gray line is the fundamental relation from \citet[their figure 1]{2006AJ....131..185R}.}

    \label{fig:FP1}
\end{figure} 
The fundamental plane links the structural parameters of ETGs, specifically their effective radii ($R_{\rm eff}$), surface brightnesses, and velocity dispersions. This relation was first introduced by \citet{1987ApJ...313...42D} and subsequently refined by \citet{1992ApJ...399..462B}. It serves as a powerful tool for constraining galaxy formation and evolution models, offering insights into the underlying physical processes that shape the structural properties of galaxies. The fundamental plane reflects the dynamical equilibrium between gravitational forces and internal stellar motions, making it essential for understanding the mass assembly history and stellar population properties of ETGs. By examining deviations from the fundamental plane, clues about the formation mechanisms, merger histories, and environmental influences that shape these galaxies can be uncovered.

The Kormendy relation (\citealt{1977ApJ...218..333K}) establishes a connection between the surface brightnesses and the effective radii of galaxies, offering valuable insights into their structural properties and formation histories. This empirical relation provides a means to study the size-luminosity distribution of galaxies and their dependence on various physical parameters. By examining the slope and scatter of the Kormendy relation, researchers can infer information about the stellar populations, mass distributions, and dynamical states of galaxies. Deviations from the Kormendy relation may indicate the presence of distinct morphological components, such as bulges, disks, or nuclear star clusters, offering clues about the assembly histories and evolutionary pathways of galaxies. Understanding the Kormendy relation is crucial for unraveling the complex interplay of physical processes that govern galaxy formation and evolution.

Figure \ref{fig:FP1} shows that both the merged models and the model relics conform to the observed relation for early-type ellipticals. However, there are notable differences: the merged models exhibit much larger effective radii ($R_{\rm eff}$), and they also have higher velocity dispersions compared to their model relics which puts the model relics on the lower-left end of the observed relation and merged models on the observed relation where most of the observed galaxies reside. This difference is due to the compact nature of the model relics.

\begin{figure}
\includegraphics[width=10.5 cm]{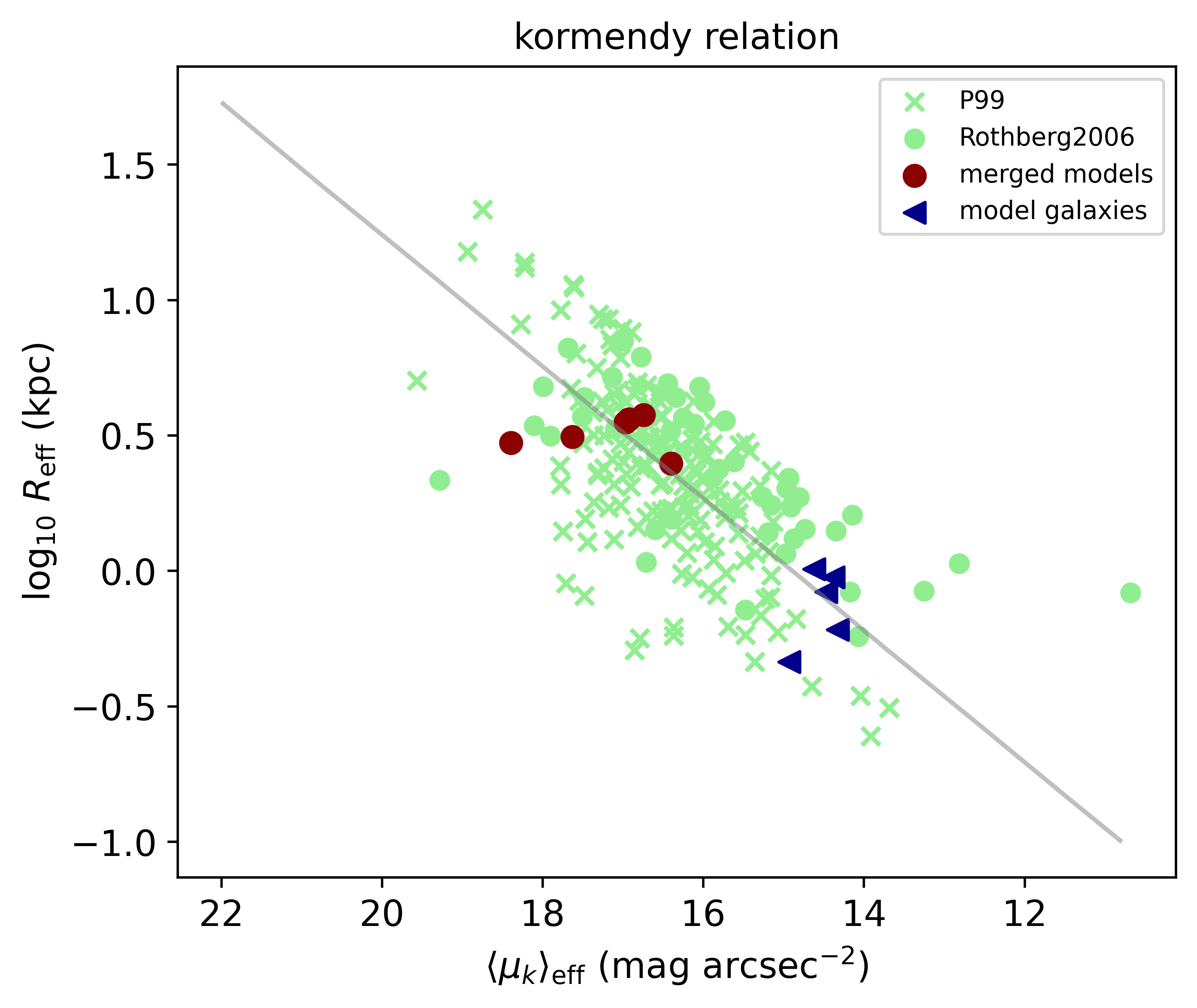}
    \caption{K-band Kormendy relation between $R_{\rm eff}$ and $\langle\mu_{K}\rangle_{\rm eff}$ of elliptical galaxies. The symbols are as in Fig. \ref{fig:FP1} and the gray line is the best - fit to the ETGs from \citet[their equation 14]{1998AJ....116.1606P} as reported in \citet{2006AJ....131..185R}.}

    \label{fig:kormendy}
\end{figure} 

This relationship is further illustrated in Figure \ref{fig:kormendy}, which displays the Kormendy relation. The data points in this figure resemble those in Figure \ref{fig:FP1}. In the Kormendy relation plot, the merged models are situated in the upper left region among the observed galaxies. In contrast, the model relics are located in the bottom right region, where the elliptical relation is less distinct. This positioning highlights the significant impact of the compactness of the model relics. Their compact nature makes them stand out as unique objects compared to typical ellipticals and discs.

\subsection{Effective phase-space density}

\begin{figure}
\includegraphics[width=10.5 cm]{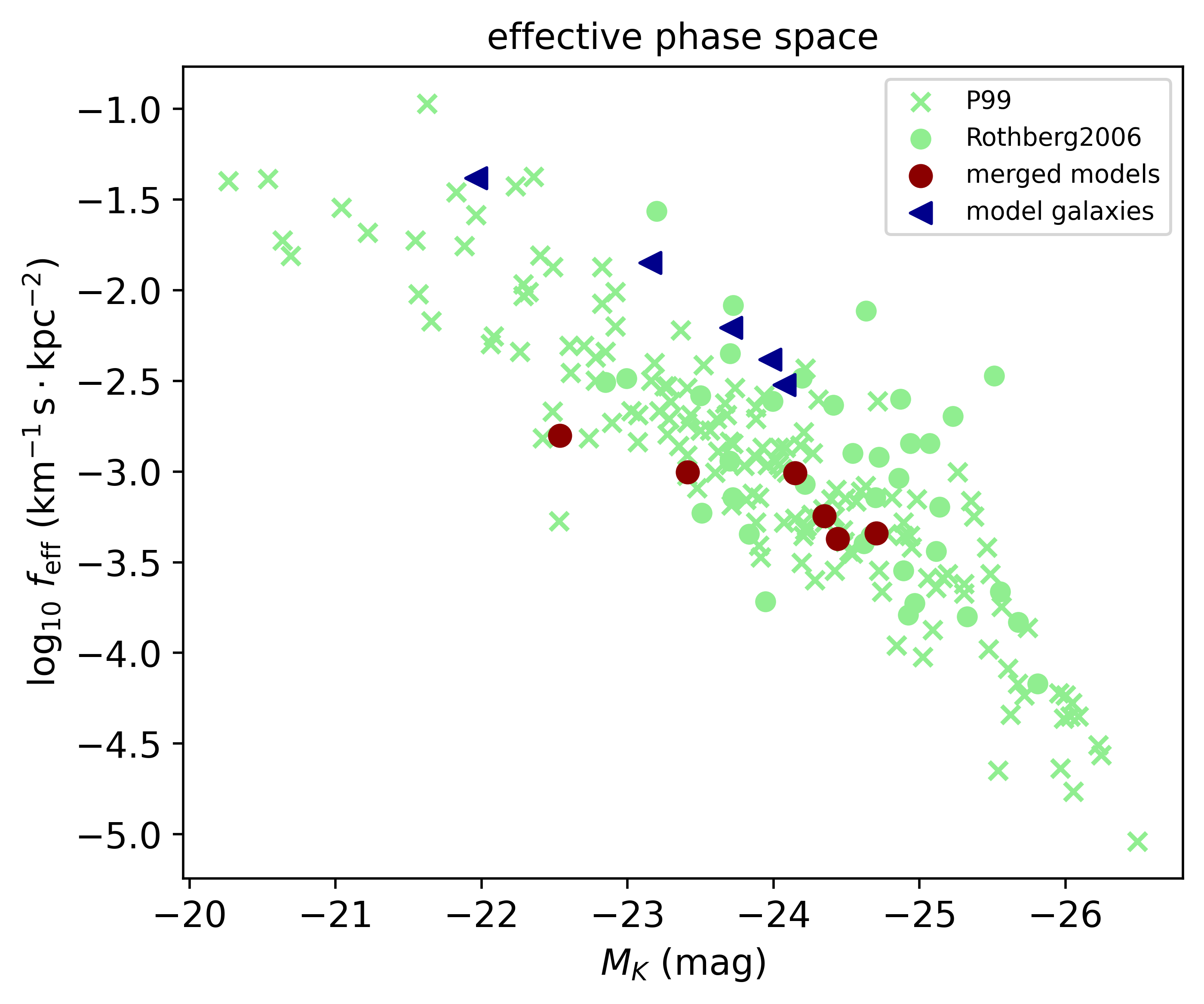}
    \caption{Effective phase space density ($f_{\rm eff}$) relation with absolute magnitude $M_{\rm K}$. The symbols are as in Fig. \ref{fig:FP1}.}

    \label{fig:feff}
\end{figure}

Another observational relation that can be studied is the effective-phase-space density ($f_{\rm eff}$) which is defined as (see \citealt{2006AJ....131..185R}),
\begin{equation}
    f_{\rm eff} = \frac{1}{V_{\sigma} \ R_{\rm eff}^{2}} \ \ (\text{km}^{-1} \text{s} \cdot \text{kpc}^{-2})
\end{equation}
The \( f_{\rm eff} \) serves as an observational proxy for the true six-dimensional phase-space density, which cannot be directly measured from projected observational data. Here, \( V_{\sigma} \) represents the velocity dispersion, encapsulating the spread of stellar velocities, while \( R_{\rm eff} \) defines the characteristic spatial scale of the galaxy. This formulation captures the general behavior of phase-space density, which is expected to decrease during mergers due to the increase in both velocity dispersion and galaxy size from violent relaxation. In Figure \ref{fig:feff}, the effective-phase-space density is plotted against the absolute magnitude ($M_{\rm K}$, Equation \ref{eq:absolutemagnitude}). The merged models are comparable to the observed elliptical galaxies, while the model relics have effective-phase space densities a factor of 4 to 10 higher than the merged galaxies.

The above correlations suggest that the evolutionary pathways of galaxies significantly influence their observed characteristics, with monolithic collapse leading to more compact galaxies in the early Universe, while later merger-driven formation processes of mostly finished elliptical galaxies result in less compact systems. This interpretation can be seen to align with the hierarchical model of galaxy formation, where early universe conditions and subsequent evolutionary processes both play crucial roles in determining the final morphology of galaxies. The mergers between elliptical galaxies must have occurred during the first Gyr after their formation. Mergers between red massive galaxies are observed to have been rare at redshift z $<$ 0.3 \citep{2008ApJ...679..260M}, being supported by the number of elliptical galaxies not changing significantly since 6 Gyr ago \citep{2010A&A...509A..78D}. In a MOND-based cosmology the dominance of the galaxy population through thin disk galaxies may arise if disk galaxy seeds form from the collapse of rotating pre-galactic gas clouds that continue to grow through near-constant gas accretion \citep{2020MNRAS.498.5652K}. In overdense regions that evolve to galaxy groups or clusters, pre-galactic gas clouds that are negligibly rotating collapsed to form one to a few ETG that can merge.

\section{Conclusions}
\label{sec:conclusion}

Recent results from the James Webb Space Telescope (\citealt{2024arXiv240416963X}) reveal that the massive galaxy GN-z11 at z=10.6 possesses a rotating disc, indicating a weak feedback mechanism in its formation. This observation is consistent with MOND predictions of ETG formation via the monolithic collapse of primordial gas clouds (\citealt{2022MNRAS.516.1081E}). This suggests that ETGs may initially form as rotating compact stellar discs and later assume an elliptical morphology following a few merger events, without altering intrinsic properties such as star-formation timescales. Such merger events would occur if multiple ETGs form at the centre of galaxy groups and clusters.

In this study, we demonstrate that relics formed through the monolithic collapse of primordial gas clouds align closely with the observed ETG relations at the faint and rotational end (Section \ref{sec:results}). Galaxies resulting from the merging of two relics exhibit properties that also align with observed ETGs, particularly in terms of their predominant elliptical shapes and less rotational properties. This suggests that the elliptical galaxies we observe may indeed be the results of such merging events, although \citet{2008A&A...486..763P} using their chemical evolution models find that dry mergers of dwarf galaxies alone cannot form massive ETGs. Conversely, most of the compact relic massive galaxies that do not conform to the conventional understanding of ETGs—those observed with a rotating disk rather than being solely dispersion-dominated, as noted by \citet{2024arXiv240416963X}—may be formed through monolithic collapse in MOND as the non-merger models studied here (\citealt{2024MNRAS.528.4264E}). A comprehensive study of early structure formation in MOND-based cosmology \citep{2024Univ...10...48M} could potentially elucidate the true nature of these ETGs.

The findings of \citet{2011MNRAS.414.2923K} show that alignment between isophotal and kinematic axes in most observed ETGs supports a simpler, two-galaxy merger scenario. Complex merger histories, as seen in more massive ETGs, likely involve additional relics, leading to axis misalignments and twisted isophotes. Our results suggest that a MOND-based cosmological model might naturally produce such variations, with more massive ETGs exhibiting complex dynamics, while less massive ones align more readily. Although a full exploration of these scenarios is beyond our current computational capacity, this work lays important groundwork for future MOND cosmological studies to examine how ETG properties might emerge from these dynamics.

We would like to note that while a comparison between simulated galaxies and observed galaxies of similar masses would be ideal, this approach presents challenges. The masses of the observed galaxies are not explicitly provided in the literature and would need to be estimated indirectly from luminosity data. This process involves additional assumptions and potential uncertainties, which could affect the accuracy of such a comparison especially in view of the variation of the galaxy-wide stellar initial mass function \citep{2024arXiv241007311K}. Consequently, we have opted to conduct a broader comparison across the sample but will explore the use of mass estimates where feasible in future studies.

\citet{2017MNRAS.468.4216Y} describe a population of local compact elliptical galaxies that obey the mass-size relation of their high-redshift counterparts, while most ETGs display an increase in their radii with redshift. This increase can be attributed to mergers, as studied here, and the local compact elliptical galaxies would be our relic models that did not merge.  In the future we will consider the properties of ETGs formed from more complex merger scenarios, with the conditions that they need to obey the downsizing constraints, and will discuss the local compact elliptical galaxies in context. The ratio between compact and more normal elliptical galaxies at a given galaxy mass will provide constraints on any possible MOND cosmological model as this ratio informs us which fraction of pre-galactic gas clouds lead to central galaxies that did not undergo mergers, i.e., formed as the only relic without a companion that later merges with it.
 
The initial exploration of structure formation in MOND cosmology was conducted by \citet{1998MNRAS.296.1009S} who proposed that the thermal and dynamical history of the early MOND Universe mirrors that of the standard Big Bang model, with predictions relevant to the nucleosynthesis of light elements remaining applicable after non-relativistic matter dominates the mass density of the Universe. Subsequent cosmological structure formation simulations in the $\nu$HDM MOND model \citep{2013ApJ...772...10K, 2023MNRAS.523..453W} have shown that the formation of a cosmic web of filaments and voids is not unique to standard Einstein/Newton-based cosmology, indicating that MOND-based cosmological frameworks can also account for this key feature of the universe's large-scale structure. However, it should be noted that a MOND cosmological model does not automatically resolve all issues, as exemplified by the $\nu$HDM model, which appears to form galaxies too late and results in overly massive structures overall \citep{2023arXiv230911552K}.


\vspace{6pt} 




\authorcontributions{R.E performed the calculations and data analysis. P.K suggested the project idea. R.E and P.K developed the manuscript.}

\acknowledgments{The author R.E would like to thank KFPP for their support. P.K would like to thank the DAAD Bonn - Eastern European Exchange programme at the University of Bonn and Prague for support.}

\begin{adjustwidth}{-\extralength}{0cm}

\reftitle{References}

\PublishersNote{}
\end{adjustwidth}

\begin{thebibliography}{999}

\bibitem[{Kormendy} et~al.(2009){Kormendy}, {Fisher}, {Cornell}, and
  {Bender}]{2009ApJS..182..216K}
{Kormendy}, J.; {Fisher}, D.B.; {Cornell}, M.E.; {Bender}, R.
\newblock {Structure and Formation of Elliptical and Spheroidal Galaxies}.
{\bf 2009}, {\em 182},~216--309,
  \href{http://arxiv.org/abs/0810.1681}{{\normalfont
  [arXiv:astro-ph/0810.1681]}}.
\newblock {\url{https://doi.org/10.1088/0067-0049/182/1/216}}.

\bibitem[{Emsellem} et~al.(2011){Emsellem}, {Cappellari}, {Krajnovi{\'c}},
  {Alatalo}, {Blitz}, {Bois}, {Bournaud}, {Bureau}, {Davies}, {Davis}, {de
  Zeeuw}, {Khochfar}, {Kuntschner}, {Lablanche}, {McDermid}, {Morganti},
  {Naab}, {Oosterloo}, {Sarzi}, {Scott}, {Serra}, {van de Ven}, {Weijmans}, and
  {Young}]{2011MNRAS.414..888E}
{Emsellem}, E.; {Cappellari}, M.; {Krajnovi{\'c}}, D.; {Alatalo}, K.; {Blitz},
  L.; {Bois}, M.; {Bournaud}, F.; {Bureau}, M.; {Davies}, R.L.; {Davis}, T.A.;
  et~al.
\newblock {The ATLAS$^{3D}$ project - III. A census of the stellar angular
  momentum within the effective radius of early-type galaxies: unveiling the
  distribution of fast and slow rotators}.
{\bf 2011}, {\em 414},~888--912,
  \href{http://arxiv.org/abs/1102.4444}{{\normalfont
  [arXiv:astro-ph.CO/1102.4444]}}.
\newblock {\url{https://doi.org/10.1111/j.1365-2966.2011.18496.x}}.

\bibitem[{Cappellari} et~al.(2011){Cappellari}, {Emsellem}, {Krajnovi{\'c}},
  {McDermid}, {Scott}, {Verdoes Kleijn}, {Young}, {Alatalo}, {Bacon}, {Blitz},
  {Bois}, {Bournaud}, {Bureau}, {Davies}, {Davis}, {de Zeeuw}, {Duc},
  {Khochfar}, {Kuntschner}, {Lablanche}, {Morganti}, {Naab}, {Oosterloo},
  {Sarzi}, {Serra}, and {Weijmans}]{2011MNRAS.413..813C}
{Cappellari}, M.; {Emsellem}, E.; {Krajnovi{\'c}}, D.; {McDermid}, R.M.;
  {Scott}, N.; {Verdoes Kleijn}, G.A.; {Young}, L.M.; {Alatalo}, K.; {Bacon},
  R.; {Blitz}, L.;  et~al.
\newblock {The ATLAS$^{3D}$ project - I. A volume-limited sample of 260 nearby
  early-type galaxies: science goals and selection criteria}.
{\bf 2011}, {\em 413},~813--836,
  \href{http://arxiv.org/abs/1012.1551}{{\normalfont
  [arXiv:astro-ph.CO/1012.1551]}}.
\newblock {\url{https://doi.org/10.1111/j.1365-2966.2010.18174.x}}.

\bibitem[{Dressler}(1980)]{1980ApJ...236..351D}
{Dressler}, A.
\newblock {Galaxy morphology in rich clusters: implications for the formation
  and evolution of galaxies.}
{\bf 1980}, {\em 236},~351--365.
\newblock {\url{https://doi.org/10.1086/157753}}.

\bibitem[{Whitmore} et~al.(1993){Whitmore}, {Gilmore}, and
  {Jones}]{1993ApJ...407..489W}
{Whitmore}, B.C.; {Gilmore}, D.M.; {Jones}, C.
\newblock {What Determines the Morphological Fractions in Clusters of
  Galaxies?}
 {\bf 1993}, {\em 407},~489.
\newblock {\url{https://doi.org/10.1086/172531}}.

\bibitem[{Trujillo} et~al.(2007){Trujillo}, {Conselice}, {Bundy}, {Cooper},
  {Eisenhardt}, and {Ellis}]{2007MNRAS.382..109T}
{Trujillo}, I.; {Conselice}, C.J.; {Bundy}, K.; {Cooper}, M.C.; {Eisenhardt},
  P.; {Ellis}, R.S.
\newblock {Strong size evolution of the most massive galaxies since z
  {\ensuremath{\sim}} 2}.
{\bf 2007}, {\em 382},~109--120,
  \href{http://arxiv.org/abs/0709.0621}{{\normalfont
  [arXiv:astro-ph/0709.0621]}}.
\newblock {\url{https://doi.org/10.1111/j.1365-2966.2007.12388.x}}.

\bibitem[{Kriek} et~al.(2009){Kriek}, {van Dokkum}, {Labb{\'e}}, {Franx},
  {Illingworth}, {Marchesini}, and {Quadri}]{2009ApJ...700..221K}
{Kriek}, M.; {van Dokkum}, P.G.; {Labb{\'e}}, I.; {Franx}, M.; {Illingworth},
  G.D.; {Marchesini}, D.; {Quadri}, R.F.
\newblock {An Ultra-Deep Near-Infrared Spectrum of a Compact Quiescent Galaxy
  at z = 2.2}.
{\bf 2009}, {\em 700},~221--231,
  \href{http://arxiv.org/abs/0905.1692}{{\normalfont
  [arXiv:astro-ph.CO/0905.1692]}}.
\newblock {\url{https://doi.org/10.1088/0004-637X/700/1/221}}.

\bibitem[{Haslbauer} et~al.(2022){Haslbauer}, {Kroupa}, {Zonoozi}, and
  {Haghi}]{2022ApJ...939L..31H}
{Haslbauer}, M.; {Kroupa}, P.; {Zonoozi}, A.H.; {Haghi}, H.
\newblock {Has JWST Already Falsified Dark-matter-driven Galaxy Formation?}
 {\bf 2022}, {\em 939},~L31,
  \href{http://arxiv.org/abs/2210.14915}{{\normalfont
  [arXiv:astro-ph.GA/2210.14915]}}.
\newblock {\url{https://doi.org/10.3847/2041-8213/ac9a50}}.

\bibitem[{Donnan} et~al.(2023){Donnan}, {McLeod}, {Dunlop}, {McLure},
  {Carnall}, {Begley}, {Cullen}, {Hamadouche}, {Bowler}, {Magee}, {McCracken},
  {Milvang-Jensen}, {Moneti}, and {Targett}]{2023MNRAS.518.6011D}
{Donnan}, C.T.; {McLeod}, D.J.; {Dunlop}, J.S.; {McLure}, R.J.; {Carnall},
  A.C.; {Begley}, R.; {Cullen}, F.; {Hamadouche}, M.L.; {Bowler}, R.A.A.;
  {Magee}, D.;  et~al.
\newblock {The evolution of the galaxy UV luminosity function at redshifts z
  $\simeq$ 8 - 15 from deep JWST and ground-based near-infrared imaging}.
 {\bf 2023}, {\em 518},~6011--6040,
  \href{http://arxiv.org/abs/2207.12356}{{\normalfont
  [arXiv:astro-ph.GA/2207.12356]}}.
\newblock {\url{https://doi.org/10.1093/mnras/stac3472}}.

\bibitem[{Harikane} et~al.(2023){Harikane}, {Ouchi}, {Oguri}, {Ono},
  {Nakajima}, {Isobe}, {Umeda}, {Mawatari}, and {Zhang}]{2023ApJS..265....5H}
{Harikane}, Y.; {Ouchi}, M.; {Oguri}, M.; {Ono}, Y.; {Nakajima}, K.; {Isobe},
  Y.; {Umeda}, H.; {Mawatari}, K.; {Zhang}, Y.
\newblock {A Comprehensive Study of Galaxies at z 9-16 Found in the Early JWST
  Data: Ultraviolet Luminosity Functions and Cosmic Star Formation History at
  the Pre-reionization Epoch}.
{\bf 2023}, {\em 265},~5,
  \href{http://arxiv.org/abs/2208.01612}{{\normalfont
  [arXiv:astro-ph.GA/2208.01612]}}.
\newblock {\url{https://doi.org/10.3847/1538-4365/acaaa9}}.

\bibitem[{Harikane} et~al.(2024){Harikane}, {Nakajima}, {Ouchi}, {Umeda},
  {Isobe}, {Ono}, {Xu}, and {Zhang}]{2024ApJ...960...56H}
{Harikane}, Y.; {Nakajima}, K.; {Ouchi}, M.; {Umeda}, H.; {Isobe}, Y.; {Ono},
  Y.; {Xu}, Y.; {Zhang}, Y.
\newblock {Pure Spectroscopic Constraints on UV Luminosity Functions and Cosmic
  Star Formation History from 25 Galaxies at z $_{spec}$ = 8.61-13.20 Confirmed
  with JWST/NIRSpec}.
 {\bf 2024}, {\em 960},~56,
  \href{http://arxiv.org/abs/2304.06658}{{\normalfont
  [arXiv:astro-ph.GA/2304.06658]}}.
\newblock {\url{https://doi.org/10.3847/1538-4357/ad0b7e}}.

\bibitem[{Djorgovski} and {Davis}(1987)]{1987ApJ...313...59D}
{Djorgovski}, S.; {Davis}, M.
\newblock {Fundamental Properties of Elliptical Galaxies}.
{\bf 1987}, {\em 313},~59.
\newblock {\url{https://doi.org/10.1086/164948}}.

\bibitem[{Binney}(1978)]{1978MNRAS.183..501B}
{Binney}, J.
\newblock {On the rotation of elliptical galaxies.}
{\bf 1978}, {\em 183},~501--514.
\newblock {\url{https://doi.org/10.1093/mnras/183.3.501}}.

\bibitem[{Kormendy}(1977)]{1977ApJ...218..333K}
{Kormendy}, J.
\newblock {Brightness distributions in compact and normal galaxies. II.
  Structure parameters of the spheroidal component.}
{\bf 1977}, {\em 218},~333--346.
\newblock {\url{https://doi.org/10.1086/155687}}.

\bibitem[{Springel} et~al.(2005){Springel}, {White}, {Jenkins}, {Frenk},
  {Yoshida}, {Gao}, {Navarro}, {Thacker}, {Croton}, {Helly}, {Peacock}, {Cole},
  {Thomas}, {Couchman}, {Evrard}, {Colberg}, and {Pearce}]{2005Natur.435..629S}
{Springel}, V.; {White}, S.D.M.; {Jenkins}, A.; {Frenk}, C.S.; {Yoshida}, N.;
  {Gao}, L.; {Navarro}, J.; {Thacker}, R.; {Croton}, D.; {Helly}, J.;  et~al.
\newblock {Simulations of the formation, evolution and clustering of galaxies
  and quasars}.
 {\bf 2005}, {\em 435},~629--636,
  \href{http://arxiv.org/abs/astro-ph/0504097}{{\normalfont
  [arXiv:astro-ph/astro-ph/0504097]}}.
\newblock {\url{https://doi.org/10.1038/nature03597}}.

\bibitem[{Schaye} et~al.(2015){Schaye}, {Crain}, {Bower}, {Furlong},
  {Schaller}, {Theuns}, {Dalla Vecchia}, {Frenk}, {McCarthy}, {Helly},
  {Jenkins}, {Rosas-Guevara}, {White}, {Baes}, {Booth}, {Camps}, {Navarro},
  {Qu}, {Rahmati}, {Sawala}, {Thomas}, and {Trayford}]{2015MNRAS.446..521S}
{Schaye}, J.; {Crain}, R.A.; {Bower}, R.G.; {Furlong}, M.; {Schaller}, M.;
  {Theuns}, T.; {Dalla Vecchia}, C.; {Frenk}, C.S.; {McCarthy}, I.G.; {Helly},
  J.C.;  et~al.
\newblock {The EAGLE project: simulating the evolution and assembly of galaxies
  and their environments}.
 {\bf 2015}, {\em 446},~521--554,
  \href{http://arxiv.org/abs/1407.7040}{{\normalfont
  [arXiv:astro-ph.GA/1407.7040]}}.
\newblock {\url{https://doi.org/10.1093/mnras/stu2058}}.

\bibitem[{Pillepich} et~al.(2018){Pillepich}, {Springel}, {Nelson}, {Genel},
  {Naiman}, {Pakmor}, {Hernquist}, {Torrey}, {Vogelsberger}, {Weinberger}, and
  {Marinacci}]{Pillepich_2018}
{Pillepich}, A.; {Springel}, V.; {Nelson}, D.; {Genel}, S.; {Naiman}, J.;
  {Pakmor}, R.; {Hernquist}, L.; {Torrey}, P.; {Vogelsberger}, M.;
  {Weinberger}, R.;  et~al.
\newblock {Simulating galaxy formation with the IllustrisTNG model}.
{\bf 2018}, {\em 473},~4077--4106,
  \href{http://arxiv.org/abs/1703.02970}{{\normalfont
  [arXiv:astro-ph.GA/1703.02970]}}.
\newblock {\url{https://doi.org/10.1093/mnras/stx2656}}.

\bibitem[{Kroupa}(2012)]{2012PASA...29..395K}
{Kroupa}, P.
\newblock {The Dark Matter Crisis: Falsification of the Current Standard Model
  of Cosmology}.
{\bf 2012}, {\em 29},~395--433,
  \href{http://arxiv.org/abs/1204.2546}{{\normalfont
  [arXiv:astro-ph.CO/1204.2546]}}.
\newblock {\url{https://doi.org/10.1071/AS12005}}.

\bibitem[{Kroupa}(2015)]{2015CaJPh..93..169K}
{Kroupa}, P.
\newblock {Galaxies as simple dynamical systems: observational data disfavor
  dark matter and stochastic star formation}.
\newblock {\em Canadian Journal of Physics} {\bf 2015}, {\em 93},~169--202,
  \href{http://arxiv.org/abs/1406.4860}{{\normalfont
  [arXiv:astro-ph.GA/1406.4860]}}.
\newblock {\url{https://doi.org/10.1139/cjp-2014-0179}}.

\bibitem[{Banik} and {Zhao}(2022)]{2022Symm...14.1331B}
{Banik}, I.; {Zhao}, H.
\newblock {From Galactic Bars to the Hubble Tension: Weighing Up the
  Astrophysical Evidence for Milgromian Gravity}.
\newblock {\em Symmetry} {\bf 2022}, {\em 14},~1331,
  \href{http://arxiv.org/abs/2110.06936}{{\normalfont
  [arXiv:astro-ph.CO/2110.06936]}}.
\newblock {\url{https://doi.org/10.3390/sym14071331}}.

\bibitem[{Melia}(2022)]{2022PASP..134l1001M}
{Melia}, F.
\newblock {A Candid Assessment of Standard Cosmology}.
 {\bf 2022}, {\em 134},~121001.
\newblock {\url{https://doi.org/10.1088/1538-3873/aca51f}}.

\bibitem[{Kroupa} et~al.(2023){Kroupa}, {Gjergo}, {Asencio}, {Haslbauer},
  {Pflamm-Altenburg}, {Wittenburg}, {Samaras}, {Thies}, and
  {Oehm}]{2023arXiv230911552K}
{Kroupa}, P.; {Gjergo}, E.; {Asencio}, E.; {Haslbauer}, M.; {Pflamm-Altenburg},
  J.; {Wittenburg}, N.; {Samaras}, N.; {Thies}, I.; {Oehm}, W.
\newblock {The many tensions with dark-matter based models and implications on
  the nature of the Universe}.
\newblock {\em arXiv e-prints} {\bf 2023}, p. arXiv:2309.11552,
  \href{http://arxiv.org/abs/2309.11552}{{\normalfont
  [arXiv:astro-ph.CO/2309.11552]}}.
\newblock {\url{https://doi.org/10.48550/arXiv.2309.11552}}.

\bibitem[{Eappen} et~al.(2022){Eappen}, {Kroupa}, {Wittenburg}, {Haslbauer},
  and {Famaey}]{2022MNRAS.516.1081E}
{Eappen}, R.; {Kroupa}, P.; {Wittenburg}, N.; {Haslbauer}, M.; {Famaey}, B.
\newblock {The formation of early-type galaxies through monolithic collapse of
  gas clouds in Milgromian gravity}.
{\bf 2022}, {\em 516},~1081--1093,
  \href{http://arxiv.org/abs/2209.00024}{{\normalfont
  [arXiv:astro-ph.GA/2209.00024]}}.
\newblock {\url{https://doi.org/10.1093/mnras/stac2229}}.

\bibitem[{Katz} et~al.(2013){Katz}, {McGaugh}, {Teuben}, and
  {Angus}]{2013ApJ...772...10K}
{Katz}, H.; {McGaugh}, S.; {Teuben}, P.; {Angus}, G.W.
\newblock {Galaxy Cluster Bulk Flows and Collision Velocities in QUMOND}.
{\bf 2013}, {\em 772},~10,
  \href{http://arxiv.org/abs/1305.3651}{{\normalfont
  [arXiv:astro-ph.CO/1305.3651]}}.
\newblock {\url{https://doi.org/10.1088/0004-637X/772/1/10}}.

\bibitem[{Wittenburg} et~al.(2023){Wittenburg}, {Kroupa}, {Banik}, {Candlish},
  and {Samaras}]{2023MNRAS.523..453W}
{Wittenburg}, N.; {Kroupa}, P.; {Banik}, I.; {Candlish}, G.; {Samaras}, N.
\newblock {Hydrodynamical structure formation in Milgromian cosmology}.
{\bf 2023}, {\em 523},~453--473,
  \href{http://arxiv.org/abs/2305.05696}{{\normalfont
  [arXiv:astro-ph.CO/2305.05696]}}.
\newblock {\url{https://doi.org/10.1093/mnras/stad1371}}.

\bibitem[{McGaugh}(2024)]{2024Univ...10...48M}
{McGaugh}, S.S.
\newblock {Discord in Concordance Cosmology and Anomalously Massive Early
  Galaxies}.
\newblock {\em Universe} {\bf 2024}, {\em 10},~48,
  \href{http://arxiv.org/abs/2312.03127}{{\normalfont
  [arXiv:astro-ph.CO/2312.03127]}}.
\newblock {\url{https://doi.org/10.3390/universe10010048}}.

\bibitem[{Milgrom}(1983)]{1983ApJ...270..365M}
{Milgrom}, M.
\newblock {A modification of the Newtonian dynamics as a possible alternative
  to the hidden mass hypothesis.}
 {\bf 1983}, {\em 270},~365--370.
\newblock {\url{https://doi.org/10.1086/161130}}.

\bibitem[{Famaey} and {McGaugh}(2012)]{2012LRR....15...10F}
{Famaey}, B.; {McGaugh}, S.S.
\newblock {Modified Newtonian Dynamics (MOND): Observational Phenomenology and
  Relativistic Extensions}.
\newblock {\em Living Reviews in Relativity} {\bf 2012}, {\em 15},~10,
  \href{http://arxiv.org/abs/1112.3960}{{\normalfont
  [arXiv:astro-ph.CO/1112.3960]}}.
\newblock {\url{https://doi.org/10.12942/lrr-2012-10}}.

\bibitem[{Famaey} and {Durakovic}(2025)]{2025arXiv250117006F}
{Famaey}, B.; {Durakovic}, A.
\newblock {Modified Newtonian Dynamics (MOND)}.
\newblock {\em arXiv e-prints} {\bf 2025}, p. arXiv:2501.17006,
  \href{http://arxiv.org/abs/2501.17006}{{\normalfont
  [arXiv:astro-ph.GA/2501.17006]}}.
\newblock {\url{https://doi.org/10.48550/arXiv.2501.17006}}.

\bibitem[{Bekenstein} and {Milgrom}(1984)]{1984ApJ...286....7B}
{Bekenstein}, J.; {Milgrom}, M.
\newblock {Does the missing mass problem signal the breakdown of Newtonian
  gravity?}
 {\bf 1984}, {\em 286},~7--14.
\newblock {\url{https://doi.org/10.1086/162570}}.

\bibitem[{Milgrom}(2010)]{2010MNRAS.403..886M}
{Milgrom}, M.
\newblock {Quasi-linear formulation of MOND}.
{\bf 2010}, {\em 403},~886--895,
  \href{http://arxiv.org/abs/0911.5464}{{\normalfont
  [arXiv:astro-ph.CO/0911.5464]}}.
\newblock {\url{https://doi.org/10.1111/j.1365-2966.2009.16184.x}}.

\bibitem[{Milgrom}(2008)]{2008arXiv0801.3133M}
{Milgrom}, M.
\newblock {The MOND paradigm}.
\newblock {\em arXiv e-prints} {\bf 2008}, p. arXiv:0801.3133,
  \href{http://arxiv.org/abs/0801.3133}{{\normalfont
  [arXiv:astro-ph/0801.3133]}}.

\bibitem[{Milgrom}(2014)]{2014SchpJ...931410M}
{Milgrom}, M.
\newblock {The MOND paradigm of modified dynamics}.
\newblock {\em Scholarpedia} {\bf 2014}, {\em 9},~31410.
\newblock {\url{https://doi.org/10.4249/scholarpedia.31410}}.

\bibitem[{McGaugh} et~al.(2000){McGaugh}, {Schombert}, {Bothun}, and {de
  Blok}]{2000ApJ...533L..99M}
{McGaugh}, S.S.; {Schombert}, J.M.; {Bothun}, G.D.; {de Blok}, W.J.G.
\newblock {The Baryonic Tully-Fisher Relation}.
{\bf 2000}, {\em 533},~L99--L102,
  \href{http://arxiv.org/abs/astro-ph/0003001}{{\normalfont
  [arXiv:astro-ph/astro-ph/0003001]}}.
\newblock {\url{https://doi.org/10.1086/312628}}.

\bibitem[{McGaugh}(2005)]{2005ApJ...632..859M}
{McGaugh}, S.S.
\newblock {The Baryonic Tully-Fisher Relation of Galaxies with Extended
  Rotation Curves and the Stellar Mass of Rotating Galaxies}.
{\bf 2005}, {\em 632},~859--871,
  \href{http://arxiv.org/abs/astro-ph/0506750}{{\normalfont
  [arXiv:astro-ph/astro-ph/0506750]}}.
\newblock {\url{https://doi.org/10.1086/432968}}.

\bibitem[{McGaugh}(2012)]{2012AJ....143...40M}
{McGaugh}, S.S.
\newblock {The Baryonic Tully-Fisher Relation of Gas-rich Galaxies as a Test of
  {\ensuremath{\Lambda}}CDM and MOND}.
{\bf 2012}, {\em 143},~40,
  \href{http://arxiv.org/abs/1107.2934}{{\normalfont
  [arXiv:astro-ph.CO/1107.2934]}}.
\newblock {\url{https://doi.org/10.1088/0004-6256/143/2/40}}.

\bibitem[{Liu} et~al.(2016){Liu}, {Peng}, {Blakeslee}, {C{\^o}t{\'e}},
  {Ferrarese}, {Jord{\'a}n}, {Puzia}, {Toloba}, and
  {Zhang}]{2016ApJ...818..179L}
{Liu}, Y.; {Peng}, E.W.; {Blakeslee}, J.; {C{\^o}t{\'e}}, P.; {Ferrarese}, L.;
  {Jord{\'a}n}, A.; {Puzia}, T.H.; {Toloba}, E.; {Zhang}, H.X.
\newblock {Evidence for the Rapid Formation of Low-mass Early-type Galaxies in
  Dense Environments}.
 {\bf 2016}, {\em 818},~179,
  \href{http://arxiv.org/abs/1512.00253}{{\normalfont
  [arXiv:astro-ph.GA/1512.00253]}}.
\newblock {\url{https://doi.org/10.3847/0004-637X/818/2/179}}.

\bibitem[{Sanders}(1990)]{1990A&ARv...2....1S}
{Sanders}, R.H.
\newblock {Mass discrepancies in galaxies: dark matter and alternatives}.
{\bf 1990}, {\em 2},~1--28.
\newblock {\url{https://doi.org/10.1007/BF00873540}}.

\bibitem[{McGaugh}(2004)]{2004ApJ...609..652M}
{McGaugh}, S.S.
\newblock {The Mass Discrepancy-Acceleration Relation: Disk Mass and the Dark
  Matter Distribution}.
 {\bf 2004}, {\em 609},~652--666,
  \href{http://arxiv.org/abs/astro-ph/0403610}{{\normalfont
  [arXiv:astro-ph/astro-ph/0403610]}}.
\newblock {\url{https://doi.org/10.1086/421338}}.

\bibitem[{Lelli} et~al.(2017){Lelli}, {McGaugh}, {Schombert}, and
  {Pawlowski}]{2017ApJ...836..152L}
{Lelli}, F.; {McGaugh}, S.S.; {Schombert}, J.M.; {Pawlowski}, M.S.
\newblock {One Law to Rule Them All: The Radial Acceleration Relation of
  Galaxies}.
{\bf 2017}, {\em 836},~152,
  \href{http://arxiv.org/abs/1610.08981}{{\normalfont
  [arXiv:astro-ph.GA/1610.08981]}}.
\newblock {\url{https://doi.org/10.3847/1538-4357/836/2/152}}.

\bibitem[{L{\"u}ghausen} et~al.(2015){L{\"u}ghausen}, {Famaey}, and
  {Kroupa}]{2015CaJPh..93..232L}
{L{\"u}ghausen}, F.; {Famaey}, B.; {Kroupa}, P.
\newblock {Phantom of RAMSES (POR): A new Milgromian dynamicsN-body code}.
\newblock {\em Canadian Journal of Physics} {\bf 2015}, {\em 93},~232--241,
  \href{http://arxiv.org/abs/1405.5963}{{\normalfont
  [arXiv:astro-ph.GA/1405.5963]}}.
\newblock {\url{https://doi.org/10.1139/cjp-2014-0168}}.

\bibitem[{Nagesh} et~al.(2021){Nagesh}, {Banik}, {Thies}, {Kroupa}, {Famaey},
  {Wittenburg}, {Parziale}, and {Haslbauer}]{2021arXiv210111011N}
{Nagesh}, S.T.; {Banik}, I.; {Thies}, I.; {Kroupa}, P.; {Famaey}, B.;
  {Wittenburg}, N.; {Parziale}, R.; {Haslbauer}, M.
\newblock {The Phantom of RAMSES user guide for galaxy simulations using
  Milgromian and Newtonian gravity}.
\newblock {\em arXiv e-prints} {\bf 2021}, p. arXiv:2101.11011,
  \href{http://arxiv.org/abs/2101.11011}{{\normalfont
  [arXiv:astro-ph.IM/2101.11011]}}.

\bibitem[{Teyssier}(2002)]{2002A&A...385..337T}
{Teyssier}, R.
\newblock {Cosmological hydrodynamics with adaptive mesh refinement. A new high
  resolution code called RAMSES}.
 {\bf 2002}, {\em 385},~337--364,
  \href{http://arxiv.org/abs/astro-ph/0111367}{{\normalfont
  [arXiv:astro-ph/astro-ph/0111367]}}.
\newblock {\url{https://doi.org/10.1051/0004-6361:20011817}}.

\bibitem[{Renaud} et~al.(2016){Renaud}, {Famaey}, and
  {Kroupa}]{2016MNRAS.463.3637R}
{Renaud}, F.; {Famaey}, B.; {Kroupa}, P.
\newblock {Star formation triggered by galaxy interactions in modified
  gravity}.
{\bf 2016}, {\em 463},~3637--3652,
  \href{http://arxiv.org/abs/1609.04407}{{\normalfont
  [arXiv:astro-ph.GA/1609.04407]}}.
\newblock {\url{https://doi.org/10.1093/mnras/stw2331}}.

\bibitem[{Thomas} et~al.(2017){Thomas}, {Famaey}, {Ibata}, {L{\"u}ghausen}, and
  {Kroupa}]{2017A&A...603A..65T}
{Thomas}, G.F.; {Famaey}, B.; {Ibata}, R.; {L{\"u}ghausen}, F.; {Kroupa}, P.
\newblock {Stellar streams as gravitational experiments. I. The case of
  Sagittarius}.
{\bf 2017}, {\em 603},~A65,
  \href{http://arxiv.org/abs/1705.01552}{{\normalfont
  [arXiv:astro-ph.GA/1705.01552]}}.
\newblock {\url{https://doi.org/10.1051/0004-6361/201730531}}.

\bibitem[{B{\'\i}lek} et~al.(2018){B{\'\i}lek}, {Thies}, {Kroupa}, and
  {Famaey}]{2018A&A...614A..59B}
{B{\'\i}lek}, M.; {Thies}, I.; {Kroupa}, P.; {Famaey}, B.
\newblock {MOND simulation suggests an origin for some peculiarities in the
  Local Group}.
{\bf 2018}, {\em 614},~A59,
  \href{http://arxiv.org/abs/1712.04938}{{\normalfont
  [arXiv:astro-ph.GA/1712.04938]}}.
\newblock {\url{https://doi.org/10.1051/0004-6361/201731939}}.

\bibitem[{Banik} et~al.(2018){Banik}, {O'Ryan}, and
  {Zhao}]{2018MNRAS.477.4768B}
{Banik}, I.; {O'Ryan}, D.; {Zhao}, H.
\newblock {Origin of the Local Group satellite planes}.
{\bf 2018}, {\em 477},~4768--4791,
  \href{http://arxiv.org/abs/1802.00440}{{\normalfont
  [arXiv:astro-ph.GA/1802.00440]}}.
\newblock {\url{https://doi.org/10.1093/mnras/sty919}}.

\bibitem[{B{\'\i}lek} et~al.(2021){B{\'\i}lek}, {Thies}, {Kroupa}, and
  {Famaey}]{2021Galax...9..100B}
{B{\'\i}lek}, M.; {Thies}, I.; {Kroupa}, P.; {Famaey}, B.
\newblock {Are Disks of Satellites Comprised of Tidal Dwarf Galaxies?}
\newblock {\em Galaxies} {\bf 2021}, {\em 9},~100,
  \href{http://arxiv.org/abs/2111.05306}{{\normalfont
  [arXiv:astro-ph.GA/2111.05306]}}.
\newblock {\url{https://doi.org/10.3390/galaxies9040100}}.

\bibitem[{Thomas} et~al.(2018){Thomas}, {Famaey}, {Ibata}, {Renaud}, {Martin},
  and {Kroupa}]{2018A&A...609A..44T}
{Thomas}, G.F.; {Famaey}, B.; {Ibata}, R.; {Renaud}, F.; {Martin}, N.F.;
  {Kroupa}, P.
\newblock {Stellar streams as gravitational experiments. II. Asymmetric tails
  of globular cluster streams}.
 {\bf 2018}, {\em 609},~A44,
  \href{http://arxiv.org/abs/1709.01934}{{\normalfont
  [arXiv:astro-ph.GA/1709.01934]}}.
\newblock {\url{https://doi.org/10.1051/0004-6361/201731609}}.

\bibitem[{Kroupa} et~al.(2022){Kroupa}, {Jerabkova}, {Thies},
  {Pflamm-Altenburg}, {Famaey}, {Boffin}, {Dabringhausen}, {Beccari}, {Prusti},
  {Boily}, {Haghi}, {Wu}, {Haas}, {Zonoozi}, {Thomas}, {{\v{S}}ubr}, and
  {Aarseth}]{2022MNRAS.517.3613K}
{Kroupa}, P.; {Jerabkova}, T.; {Thies}, I.; {Pflamm-Altenburg}, J.; {Famaey},
  B.; {Boffin}, H.M.J.; {Dabringhausen}, J.; {Beccari}, G.; {Prusti}, T.;
  {Boily}, C.;  et~al.
\newblock {Asymmetrical tidal tails of open star clusters: stars crossing their
  cluster's pr{\'a}h$^{{\textdagger}}$ challenge Newtonian gravitation}.
 {\bf 2022}, {\em 517},~3613--3639,
  \href{http://arxiv.org/abs/2210.13472}{{\normalfont
  [arXiv:astro-ph.GA/2210.13472]}}.
\newblock {\url{https://doi.org/10.1093/mnras/stac2563}}.

\bibitem[{Wittenburg} et~al.(2020){Wittenburg}, {Kroupa}, and
  {Famaey}]{2020ApJ...890..173W}
{Wittenburg}, N.; {Kroupa}, P.; {Famaey}, B.
\newblock {The Formation of Exponential Disk Galaxies in MOND}.
{\bf 2020}, {\em 890},~173,
  \href{http://arxiv.org/abs/2002.01941}{{\normalfont
  [arXiv:astro-ph.GA/2002.01941]}}.
\newblock {\url{https://doi.org/10.3847/1538-4357/ab6d73}}.

\bibitem[{Banik} et~al.(2020){Banik}, {Thies}, {Famaey}, {Candlish}, {Kroupa},
  and {Ibata}]{2020ApJ...905..135B}
{Banik}, I.; {Thies}, I.; {Famaey}, B.; {Candlish}, G.; {Kroupa}, P.; {Ibata},
  R.
\newblock {The Global Stability of M33 in MOND}.
 {\bf 2020}, {\em 905},~135,
  \href{http://arxiv.org/abs/2011.12293}{{\normalfont
  [arXiv:astro-ph.GA/2011.12293]}}.
\newblock {\url{https://doi.org/10.3847/1538-4357/abc623}}.

\bibitem[{B{\'\i}lek} et~al.(2021){B{\'\i}lek}, {Zhao}, {Famaey}, {M{\"u}ller},
  {Kroupa}, and {Ibata}]{2021arXiv210705667B}
{B{\'\i}lek}, M.; {Zhao}, H.; {Famaey}, B.; {M{\"u}ller}, O.; {Kroupa}, P.;
  {Ibata}, R.
\newblock {Evolution of globular-cluster systems of ultra-diffuse galaxies due
  to dynamical friction in MOND gravity}.
\newblock {\em A\&A} {\bf 2021}, p. 537,
  \href{http://arxiv.org/abs/A31}{{\normalfont [arXiv:astro-ph.GA/A31]}}.

\bibitem[{L{\"u}ghausen} et~al.(2013){L{\"u}ghausen}, {Famaey}, {Kroupa},
  {Angus}, {Combes}, {Gentile}, {Tiret}, and {Zhao}]{2013MNRAS.432.2846L}
{L{\"u}ghausen}, F.; {Famaey}, B.; {Kroupa}, P.; {Angus}, G.; {Combes}, F.;
  {Gentile}, G.; {Tiret}, O.; {Zhao}, H.
\newblock {Polar ring galaxies as tests of gravity}.
{\bf 2013}, {\em 432},~2846--2853,
  \href{http://arxiv.org/abs/1304.4931}{{\normalfont
  [arXiv:astro-ph.GA/1304.4931]}}.
\newblock {\url{https://doi.org/10.1093/mnras/stt639}}.

\bibitem[{Nagesh} et~al.(2021){Nagesh}, {Banik}, {Thies}, {Kroupa}, {Famaey},
  {Wittenburg}, {Parziale}, and {Haslbauer}]{2021CaJPh..99..607N}
{Nagesh}, S.T.; {Banik}, I.; {Thies}, I.; {Kroupa}, P.; {Famaey}, B.;
  {Wittenburg}, N.; {Parziale}, R.; {Haslbauer}, M.
\newblock {The Phantom of RAMSES user guide for galaxy simulations using
  Milgromian and Newtonian gravity}.
\newblock {\em Canadian Journal of Physics} {\bf 2021}, {\em 99},~607--613,
  \href{http://arxiv.org/abs/2101.11011}{{\normalfont
  [arXiv:astro-ph.IM/2101.11011]}}.
\newblock {\url{https://doi.org/10.1139/cjp-2020-0624}}.

\bibitem[{Eappen} and {Kroupa}(2024)]{2024MNRAS.528.4264E}
{Eappen}, R.; {Kroupa}, P.
\newblock {The formation of compact massive relic galaxies in MOND}.
{\bf 2024}, {\em 528},~4264--4271,
  \href{http://arxiv.org/abs/2402.00103}{{\normalfont
  [arXiv:astro-ph.GA/2402.00103]}}.
\newblock {\url{https://doi.org/10.1093/mnras/stae286}}.

\bibitem[{Planck Collaboration} et~al.(2020){Planck Collaboration}, {Aghanim},
  {Akrami}, {Ashdown}, {Aumont}, {Baccigalupi}, {Ballardini}, {Banday},
  {Barreiro}, {Bartolo}, {Basak}, {Battye}, {Benabed}, {Bernard}, {Bersanelli},
  {Bielewicz}, {Bock}, {Bond}, {Borrill}, {Bouchet}, {Boulanger}, {Bucher},
  {Burigana}, {Butler}, {Calabrese}, {Cardoso}, {Carron}, {Challinor},
  {Chiang}, {Chluba}, {Colombo}, {Combet}, {Contreras}, {Crill}, {Cuttaia}, {de
  Bernardis}, {de Zotti}, {Delabrouille}, {Delouis}, {Di Valentino}, {Diego},
  {Dor{\'e}}, {Douspis}, {Ducout}, {Dupac}, {Dusini}, {Efstathiou}, {Elsner},
  {En{\ss}lin}, {Eriksen}, {Fantaye}, {Farhang}, {Fergusson},
  {Fernandez-Cobos}, {Finelli}, {Forastieri}, {Frailis}, {Fraisse},
  {Franceschi}, {Frolov}, {Galeotta}, {Galli}, {Ganga}, {G{\'e}nova-Santos},
  {Gerbino}, {Ghosh}, {Gonz{\'a}lez-Nuevo}, {G{\'o}rski}, {Gratton},
  {Gruppuso}, {Gudmundsson}, {Hamann}, {Handley}, {Hansen}, {Herranz},
  {Hildebrandt}, {Hivon}, {Huang}, {Jaffe}, {Jones}, {Karakci}, {Keih{\"a}nen},
  {Keskitalo}, {Kiiveri}, {Kim}, {Kisner}, {Knox}, {Krachmalnicoff}, {Kunz},
  {Kurki-Suonio}, {Lagache}, {Lamarre}, {Lasenby}, {Lattanzi}, {Lawrence}, {Le
  Jeune}, {Lemos}, {Lesgourgues}, {Levrier}, {Lewis}, {Liguori}, {Lilje},
  {Lilley}, {Lindholm}, {L{\'o}pez-Caniego}, {Lubin}, {Ma},
  {Mac{\'\i}as-P{\'e}rez}, {Maggio}, {Maino}, {Mandolesi}, {Mangilli},
  {Marcos-Caballero}, {Maris}, {Martin}, {Martinelli},
  {Mart{\'\i}nez-Gonz{\'a}lez}, {Matarrese}, {Mauri}, {McEwen}, {Meinhold},
  {Melchiorri}, {Mennella}, {Migliaccio}, {Millea}, {Mitra},
  {Miville-Desch{\^e}nes}, {Molinari}, {Montier}, {Morgante}, {Moss}, {Natoli},
  {N{\o}rgaard-Nielsen}, {Pagano}, {Paoletti}, {Partridge}, {Patanchon},
  {Peiris}, {Perrotta}, {Pettorino}, {Piacentini}, {Polastri}, {Polenta},
  {Puget}, {Rachen}, {Reinecke}, {Remazeilles}, {Renzi}, {Rocha}, {Rosset},
  {Roudier}, {Rubi{\~n}o-Mart{\'\i}n}, {Ruiz-Granados}, {Salvati}, {Sandri},
  {Savelainen}, {Scott}, {Shellard}, {Sirignano}, {Sirri}, {Spencer},
  {Sunyaev}, {Suur-Uski}, {Tauber}, {Tavagnacco}, {Tenti}, {Toffolatti},
  {Tomasi}, {Trombetti}, {Valenziano}, {Valiviita}, {Van Tent}, {Vibert},
  {Vielva}, {Villa}, {Vittorio}, {Wandelt}, {Wehus}, {White}, {White},
  {Zacchei}, and {Zonca}]{2020A&A...641A...6P}
{Planck Collaboration}.; {Aghanim}, N.; {Akrami}, Y.; {Ashdown}, M.; {Aumont},
  J.; {Baccigalupi}, C.; {Ballardini}, M.; {Banday}, A.J.; {Barreiro}, R.B.;
  {Bartolo}, N.;  et~al.
\newblock {Planck 2018 results. VI. Cosmological parameters}.
 {\bf 2020}, {\em 641},~A6,
  \href{http://arxiv.org/abs/1807.06209}{{\normalfont
  [arXiv:astro-ph.CO/1807.06209]}}.
\newblock {\url{https://doi.org/10.1051/0004-6361/201833910}}.

\bibitem[{Thomas} et~al.(2010){Thomas}, {Maraston}, {Schawinski}, {Sarzi}, and
  {Silk}]{2010MNRAS.404.1775T}
{Thomas}, D.; {Maraston}, C.; {Schawinski}, K.; {Sarzi}, M.; {Silk}, J.
\newblock {Environment and self-regulation in galaxy formation}.
{\bf 2010}, {\em 404},~1775--1789,
  \href{http://arxiv.org/abs/0912.0259}{{\normalfont
  [arXiv:astro-ph.CO/0912.0259]}}.
\newblock {\url{https://doi.org/10.1111/j.1365-2966.2010.16427.x}}.

\bibitem[{Bell} and {de Jong}(2001)]{2001ApJ...550..212B}
{Bell}, E.F.; {de Jong}, R.S.
\newblock {Stellar Mass-to-Light Ratios and the Tully-Fisher Relation}.
{\bf 2001}, {\em 550},~212--229,
  \href{http://arxiv.org/abs/astro-ph/0011493}{{\normalfont
  [arXiv:astro-ph/astro-ph/0011493]}}.
\newblock {\url{https://doi.org/10.1086/319728}}.

\bibitem[{Bell} et~al.(2003){Bell}, {McIntosh}, {Katz}, and
  {Weinberg}]{2003ApJS..149..289B}
{Bell}, E.F.; {McIntosh}, D.H.; {Katz}, N.; {Weinberg}, M.D.
\newblock {The Optical and Near-Infrared Properties of Galaxies. I. Luminosity
  and Stellar Mass Functions}.
 {\bf 2003}, {\em 149},~289--312,
  \href{http://arxiv.org/abs/astro-ph/0302543}{{\normalfont
  [arXiv:astro-ph/astro-ph/0302543]}}.
\newblock {\url{https://doi.org/10.1086/378847}}.

\bibitem[{Cappellari} et~al.(2006){Cappellari}, {Bacon}, {Bureau}, {Damen},
  {Davies}, {de Zeeuw}, {Emsellem}, {Falc{\'o}n-Barroso}, {Krajnovi{\'c}},
  {Kuntschner}, {McDermid}, {Peletier}, {Sarzi}, {van den Bosch}, and {van de
  Ven}]{2006MNRAS.366.1126C}
{Cappellari}, M.; {Bacon}, R.; {Bureau}, M.; {Damen}, M.C.; {Davies}, R.L.; {de
  Zeeuw}, P.T.; {Emsellem}, E.; {Falc{\'o}n-Barroso}, J.; {Krajnovi{\'c}}, D.;
  {Kuntschner}, H.;  et~al.
\newblock {The SAURON project - IV. The mass-to-light ratio, the virial mass
  estimator and the Fundamental Plane of elliptical and lenticular galaxies}.
 {\bf 2006}, {\em 366},~1126--1150,
  \href{http://arxiv.org/abs/astro-ph/0505042}{{\normalfont
  [arXiv:astro-ph/astro-ph/0505042]}}.
\newblock {\url{https://doi.org/10.1111/j.1365-2966.2005.09981.x}}.

\bibitem[{Kroupa} et~al.(2024){Kroupa}, {Gjergo}, {Jerabkova}, and
  {Yan}]{2024arXiv241007311K}
{Kroupa}, P.; {Gjergo}, E.; {Jerabkova}, T.; {Yan}, Z.
\newblock {The initial mass function of stars}.
\newblock {\em arXiv e-prints} {\bf 2024}, p. arXiv:2410.07311,
  \href{http://arxiv.org/abs/2410.07311}{{\normalfont
  [arXiv:astro-ph.GA/2410.07311]}}.
\newblock {\url{https://doi.org/10.48550/arXiv.2410.07311}}.

\bibitem[{Mo} et~al.(2010){Mo}, {van den Bosch}, and
  {White}]{2010gfe..book.....M}
{Mo}, H.; {van den Bosch}, F.C.; {White}, S.
\newblock {\em {Galaxy Formation and Evolution, Cambridge University Press}};
  2010.

\bibitem[{Binney} and {Tremaine}(2008)]{2008gady.book.....B}
{Binney}, J.; {Tremaine}, S.
\newblock {\em {Galactic Dynamics: Second Edition}};  2008.

\bibitem[{Dabringhausen} and {Fellhauer}(2016)]{2016MNRAS.460.4492D}
{Dabringhausen}, J.; {Fellhauer}, M.
\newblock {An extensive catalogue of early-type galaxies in the nearby
  Universe}.
 {\bf 2016}, {\em 460},~4492--4512,
  \href{http://arxiv.org/abs/1605.06705}{{\normalfont
  [arXiv:astro-ph.GA/1605.06705]}}.
\newblock {\url{https://doi.org/10.1093/mnras/stw1248}}.

\bibitem[{Cassata} et~al.(2011){Cassata}, {Giavalisco}, {Guo}, {Renzini},
  {Ferguson}, {Koekemoer}, {Salimbeni}, {Scarlata}, {Grogin}, {Conselice},
  {Dahlen}, {Lotz}, {Dickinson}, and {Lin}]{2011ApJ...743...96C}
{Cassata}, P.; {Giavalisco}, M.; {Guo}, Y.; {Renzini}, A.; {Ferguson}, H.;
  {Koekemoer}, A.M.; {Salimbeni}, S.; {Scarlata}, C.; {Grogin}, N.A.;
  {Conselice}, C.J.;  et~al.
\newblock {The Relative Abundance of Compact and Normal Massive Early-type
  Galaxies and Its Evolution from Redshift z \raisebox{-0.5ex}\textasciitilde 2
  to the Present}.
 {\bf 2011}, {\em 743},~96,
  \href{http://arxiv.org/abs/1106.4308}{{\normalfont
  [arXiv:astro-ph.CO/1106.4308]}}.
\newblock {\url{https://doi.org/10.1088/0004-637X/743/1/96}}.

\bibitem[{Davies} et~al.(1983){Davies}, {Efstathiou}, {Fall}, {Illingworth},
  and {Schechter}]{1983ApJ...266...41D}
{Davies}, R.L.; {Efstathiou}, G.; {Fall}, S.M.; {Illingworth}, G.; {Schechter},
  P.L.
\newblock {The kinematic properties of faint elliptical galaxies.}
{\bf 1983}, {\em 266},~41--57.
\newblock {\url{https://doi.org/10.1086/160757}}.

\bibitem[{Halliday} et~al.(2001){Halliday}, {Davies}, {Kuntschner},
  {Birkinshaw}, {Bender}, {Saglia}, and {Baggley}]{2001MNRAS.326..473H}
{Halliday}, C.; {Davies}, R.L.; {Kuntschner}, H.; {Birkinshaw}, M.; {Bender},
  R.; {Saglia}, R.P.; {Baggley}, G.
\newblock {Line-of-sight velocity distributions of low-luminosity elliptical
  galaxies}.
 {\bf 2001}, {\em 326},~473--489,
  \href{http://arxiv.org/abs/astro-ph/0103295}{{\normalfont
  [arXiv:astro-ph/astro-ph/0103295]}}.
\newblock {\url{https://doi.org/10.1046/j.1365-8711.2001.04492.x}}.

\bibitem[{Rothberg} and {Joseph}(2006)]{2006AJ....131..185R}
{Rothberg}, B.; {Joseph}, R.D.
\newblock {A Survey of Merger Remnants. II. The Emerging Kinematic and
  Photometric Correlations}.
 {\bf 2006}, {\em 131},~185--207,
  \href{http://arxiv.org/abs/astro-ph/0510019}{{\normalfont
  [arXiv:astro-ph/astro-ph/0510019]}}.
\newblock {\url{https://doi.org/10.1086/498452}}.

\bibitem[{Emsellem} et~al.(2007){Emsellem}, {Cappellari}, {Krajnovi{\'c}}, {van
  de Ven}, {Bacon}, {Bureau}, {Davies}, {de Zeeuw}, {Falc{\'o}n-Barroso},
  {Kuntschner}, {McDermid}, {Peletier}, and {Sarzi}]{2007MNRAS.379..401E}
{Emsellem}, E.; {Cappellari}, M.; {Krajnovi{\'c}}, D.; {van de Ven}, G.;
  {Bacon}, R.; {Bureau}, M.; {Davies}, R.L.; {de Zeeuw}, P.T.;
  {Falc{\'o}n-Barroso}, J.; {Kuntschner}, H.;  et~al.
\newblock {The SAURON project - IX. A kinematic classification for early-type
  galaxies}.
 {\bf 2007}, {\em 379},~401--417,
  \href{http://arxiv.org/abs/astro-ph/0703531}{{\normalfont
  [arXiv:astro-ph/astro-ph/0703531]}}.
\newblock {\url{https://doi.org/10.1111/j.1365-2966.2007.11752.x}}.

\bibitem[{Meza} et~al.(2003){Meza}, {Navarro}, {Steinmetz}, and
  {Eke}]{2003ApJ...590..619M}
{Meza}, A.; {Navarro}, J.F.; {Steinmetz}, M.; {Eke}, V.R.
\newblock {Simulations of Galaxy Formation in a {\ensuremath{\Lambda}}CDM
  Universe. III. The Dissipative Formation of an Elliptical Galaxy}.
 {\bf 2003}, {\em 590},~619--635,
  \href{http://arxiv.org/abs/astro-ph/0301224}{{\normalfont
  [arXiv:astro-ph/astro-ph/0301224]}}.
\newblock {\url{https://doi.org/10.1086/375151}}.

\bibitem[{Binney} and {Tremaine}(1987)]{1987gady.book.....B}
{Binney}, J.; {Tremaine}, S.
\newblock {\em {Galactic dynamics}};  1987.

\bibitem[{Bender} et~al.(1988){Bender}, {Doebereiner}, and
  {Moellenhoff}]{1988A&AS...74..385B}
{Bender}, R.; {Doebereiner}, S.; {Moellenhoff}, C.
\newblock {Isophote shapes of elliptical galaxies. I. The data.}
{\bf 1988}, {\em 74},~385--426.

\bibitem[{Pahre}(1999)]{1999ApJS..124..127P}
{Pahre}, M.A.
\newblock {Near-infrared Imaging of Early-Type Galaxies. II. Global Photometric
  Parameters}.
 {\bf 1999}, {\em 124},~127--169.
\newblock {\url{https://doi.org/10.1086/313249}}.

\bibitem[{Dressler} et~al.(1987){Dressler}, {Lynden-Bell}, {Burstein},
  {Davies}, {Faber}, {Terlevich}, and {Wegner}]{1987ApJ...313...42D}
{Dressler}, A.; {Lynden-Bell}, D.; {Burstein}, D.; {Davies}, R.L.; {Faber},
  S.M.; {Terlevich}, R.; {Wegner}, G.
\newblock {Spectroscopy and Photometry of Elliptical Galaxies. I. New Distance
  Estimator}.
{\bf 1987}, {\em 313},~42.
\newblock {\url{https://doi.org/10.1086/164947}}.

\bibitem[{Bender} et~al.(1992){Bender}, {Burstein}, and
  {Faber}]{1992ApJ...399..462B}
{Bender}, R.; {Burstein}, D.; {Faber}, S.M.
\newblock {Dynamically Hot Galaxies. I. Structural Properties}.
{\bf 1992}, {\em 399},~462.
\newblock {\url{https://doi.org/10.1086/171940}}.

\bibitem[{Pahre} et~al.(1998){Pahre}, {de Carvalho}, and
  {Djorgovski}]{1998AJ....116.1606P}
{Pahre}, M.A.; {de Carvalho}, R.R.; {Djorgovski}, S.G.
\newblock {Near-Infrared Imaging of Early-Type Galaxies. IV. The Physical
  Origins of the Fundamental Plane Scaling Relations}.
{\bf 1998}, {\em 116},~1606--1625,
  \href{http://arxiv.org/abs/astro-ph/9806326}{{\normalfont
  [arXiv:astro-ph/astro-ph/9806326]}}.
\newblock {\url{https://doi.org/10.1086/300545}}.

\bibitem[{Masjedi} et~al.(2008){Masjedi}, {Hogg}, and
  {Blanton}]{2008ApJ...679..260M}
{Masjedi}, M.; {Hogg}, D.W.; {Blanton}, M.R.
\newblock {The Growth of Luminous Red Galaxies by Merging}.
{\bf 2008}, {\em 679},~260--268,
  \href{http://arxiv.org/abs/0708.3240}{{\normalfont
  [arXiv:astro-ph/0708.3240]}}.
\newblock {\url{https://doi.org/10.1086/586696}}.

\bibitem[{Delgado-Serrano} et~al.(2010){Delgado-Serrano}, {Hammer}, {Yang},
  {Puech}, {Flores}, and {Rodrigues}]{2010A&A...509A..78D}
{Delgado-Serrano}, R.; {Hammer}, F.; {Yang}, Y.B.; {Puech}, M.; {Flores}, H.;
  {Rodrigues}, M.
\newblock {How was the Hubble sequence 6 Gyr ago?}
 {\bf 2010}, {\em 509},~A78,
  \href{http://arxiv.org/abs/0906.2805}{{\normalfont
  [arXiv:astro-ph.CO/0906.2805]}}.
\newblock {\url{https://doi.org/10.1051/0004-6361/200912704}}.

\bibitem[{Kroupa} et~al.(2020){Kroupa}, {Subr}, {Jerabkova}, and
  {Wang}]{2020MNRAS.498.5652K}
{Kroupa}, P.; {Subr}, L.; {Jerabkova}, T.; {Wang}, L.
\newblock {Very high redshift quasars and the rapid emergence of supermassive
  black holes}.
{\bf 2020}, {\em 498},~5652--5683,
  \href{http://arxiv.org/abs/2007.14402}{{\normalfont
  [arXiv:astro-ph.GA/2007.14402]}}.
\newblock {\url{https://doi.org/10.1093/mnras/staa2276}}.

\bibitem[{Xu} et~al.(2024){Xu}, {Ouchi}, {Yajima}, {Fukushima}, {Harikane},
  {Isobe}, {Nakajima}, {Nakane}, {Ono}, {Umeda}, {Yanagisawa}, and
  {Zhang}]{2024arXiv240416963X}
{Xu}, Y.; {Ouchi}, M.; {Yajima}, H.; {Fukushima}, H.; {Harikane}, Y.; {Isobe},
  Y.; {Nakajima}, K.; {Nakane}, M.; {Ono}, Y.; {Umeda}, H.;  et~al.
\newblock {Dynamics of GN-z11 Explored by JWST Integral Field Spectroscopy:
  Gaseous Rotating Disk at z=10.60 Suggestive of Weak Feedback?}
\newblock {\em arXiv e-prints} {\bf 2024}, p. arXiv:2404.16963,
  \href{http://arxiv.org/abs/2404.16963}{{\normalfont
  [arXiv:astro-ph.GA/2404.16963]}}.
\newblock {\url{https://doi.org/10.48550/arXiv.2404.16963}}.

\bibitem[{Pipino} and {Matteucci}(2008)]{2008A&A...486..763P}
{Pipino}, A.; {Matteucci}, F.
\newblock {Are dry mergers of ellipticals the way to reconcile model
  predictions with downsizing?}
{\bf 2008}, {\em 486},~763--769,
  \href{http://arxiv.org/abs/0805.0793}{{\normalfont
  [arXiv:astro-ph/0805.0793]}}.
\newblock {\url{https://doi.org/10.1051/0004-6361:200809395}}.

\bibitem[{Krajnovi{\'c}} et~al.(2011){Krajnovi{\'c}}, {Emsellem}, {Cappellari},
  {Alatalo}, {Blitz}, {Bois}, {Bournaud}, {Bureau}, {Davies}, {Davis}, {de
  Zeeuw}, {Khochfar}, {Kuntschner}, {Lablanche}, {McDermid}, {Morganti},
  {Naab}, {Oosterloo}, {Sarzi}, {Scott}, {Serra}, {Weijmans}, and
  {Young}]{2011MNRAS.414.2923K}
{Krajnovi{\'c}}, D.; {Emsellem}, E.; {Cappellari}, M.; {Alatalo}, K.; {Blitz},
  L.; {Bois}, M.; {Bournaud}, F.; {Bureau}, M.; {Davies}, R.L.; {Davis}, T.A.;
  et~al.
\newblock {The ATLAS$^{3D}$ project - II. Morphologies, kinemetric features and
  alignment between photometric and kinematic axes of early-type galaxies}.
{\bf 2011}, {\em 414},~2923--2949,
  \href{http://arxiv.org/abs/1102.3801}{{\normalfont
  [arXiv:astro-ph.CO/1102.3801]}}.
\newblock {\url{https://doi.org/10.1111/j.1365-2966.2011.18560.x}}.

\bibitem[{Y{\i}ld{\i}r{\i}m} et~al.(2017){Y{\i}ld{\i}r{\i}m}, {van den Bosch},
  {van de Ven}, {Mart{\'\i}n-Navarro}, {Walsh}, {Husemann}, {G{\"u}ltekin}, and
  {Gebhardt}]{2017MNRAS.468.4216Y}
{Y{\i}ld{\i}r{\i}m}, A.; {van den Bosch}, R.C.E.; {van de Ven}, G.;
  {Mart{\'\i}n-Navarro}, I.; {Walsh}, J.L.; {Husemann}, B.; {G{\"u}ltekin}, K.;
  {Gebhardt}, K.
\newblock {The structural and dynamical properties of compact elliptical
  galaxies}.
{\bf 2017}, {\em 468},~4216--4245,
  \href{http://arxiv.org/abs/1701.05898}{{\normalfont
  [arXiv:astro-ph.GA/1701.05898]}}.
\newblock {\url{https://doi.org/10.1093/mnras/stx732}}.

\bibitem[{Sanders}(1998)]{1998MNRAS.296.1009S}
{Sanders}, R.H.
\newblock {Cosmology with modified Newtonian dynamics (MOND)}.
 {\bf 1998}, {\em 296},~1009--1018,
  \href{http://arxiv.org/abs/astro-ph/9710335}{{\normalfont
  [arXiv:astro-ph/astro-ph/9710335]}}.
\newblock {\url{https://doi.org/10.1046/j.1365-8711.1998.01459.x}}.

\end{thebibliography}
\end{document}